\documentclass{aa}  
\usepackage{multirow}
\usepackage{graphicx}

\usepackage{longtable}

\usepackage{tabularx}
\usepackage{txfonts}
\begin{document}

   \title{Searching for signatures of Fe II atomic processes in spectra of active galactic nuclei}


   \author{Jelena Kova\v cevi\' c-Doj\v cinovi\'c
   \inst{1},
         Ivan Doj\v cinovi\'c
         \inst{2},
         Ma\v sa Laki\' cevi\' c
         \inst{1}
          \and
         Luka \v C. Popovi\'c
          \inst{1,3}
          }

   \institute{Astronomical Observatory, Volgina 7, 11160 Belgrade, Serbia\\    
     \email{jkovacevic@aob.bg.ac.rs}
   \and
   Faculty of Physics, University of Belgrade, Studentski Trg 12, 11000 Belgrade, Serbia\\
       \and           
       Department of Astronomy, Faculty of Mathematics, University of Belgrade, Studentski Trg 16, 11000 Belgrade, Serbia\\
       }



  \abstract
  {}
 {We use a large sample of Type 1 active galactic nuclei (AGNs) spectra in order to investigate which atomic processes are responsible for some observed properties of the Fe II emission lines and how they are connected with macroscopic physical characteristics of AGN emission regions. We especially focus on the violated relative intensities between different optical  Fe II lines, whose relative strengths do not follow the expected values according to atomic parameters. We investigated the connection between this effect and the ratio of optical to UV Fe II lines (Fe II$_{opt}$/Fe II$_{UV}$). }
{We divided the optical Fe II lines into two large line groups: consistent (Fe II$_{cons}$), whose relative intensities are in accordance with their atomic properties, and inconsistent (Fe II$_{incons}$), whose relative intensities are significantly stronger than theoretically expected. We fitted the spectra with a flexible and complex optical Fe II model, where both consistent and inconsistent Fe II lines were divided into several line groups according to their atomic characteristics and fitted independently in order to obtain more empirical clues about their properties. We focused particularly on understanding the processes that produce strong inconsistent Fe II lines, and therefore, we investigated their correlations with Fe II$_{cons}$ as well as with UV Fe II lines and some measured spectral parameters.}
{The ratios of Fe II$_{incons}$/Fe II$_{cons}$ and Fe II$_{opt}$/Fe II$_{UV}$ increase as the Eddington ratio increases and as the line widths decrease. It is possible that both ratios are affected by the process of self-absorption of stronger lines, which is responsible for the transmission of energy from the UV to the optical Fe II emission lines and, analogously, from the Fe II$_{cons}$ to the Fe II$_{incons}$ lines. In this scenario, the high Eddington ratio causes an increase in the optical depth in Fe II lines, which results in the triggering of the process of self-absorption. Measured average widths for different Fe II line groups indicate the stratification of the optical Fe II emission region. This implies that  the observed Fe II spectrum is probably a complex mixture of radiation from emission regions with different physical conditions and distances from the black hole.
}
 {}

   \keywords{galaxies: active -- galaxies: Seyfert --  galaxies: emission lines -- line: profiles -- line: formation -- atomic processes}
    \titlerunning{Searching for signatures of Fe II atomic processes in AGN spectra}
               \authorrunning{Kova\v cevi\' c-Doj\v cinovi\'c et al.}                                      

   \maketitle
%

\section{Introduction}\label{Sec1}

Iron emission lines are certainly among the most interesting features in the spectra of active galactic nuclei (AGNs) Type 1. Many of their properties, such as mechanisms of excitation, place of emission region, and some correlations with the other spectral properties, have been the subject of debate for more than 40 years \citep[for review see][]{Gaskell2022}. The Fe II ions have a complex atomic structure and a large number of possible transitions. Therefore, they emit a rich spectrum consisting of numerous iron lines that overlap and form characteristic Fe II emission bumps in the optical and UV ranges of the AGN spectra. The properties of iron lines - their appearance or non-appearance, widths, shifts, relative intensities, and variability - represent a unique laboratory for the investigation of complex atomic processes as well as the physics, kinematics, and structure of the AGN emission regions. 

The place of formation of optical Fe II lines in the AGN structure is still an open question. It has been proposed that optical Fe II lines arise at the surface of the accretion disc \citep{Collin1980, Zhang2006} in the broad line region (BLR) of AGNs where Balmer lines arise \citep{Boroson1992}, in the BLR but with a radius twice that of H$\beta$  \citep{Gaskell2007, Gaskell2009, Marinello2016}, and in an intermediate line region (ILR), which is the outermost area of the BLR, farther away from the black hole \citep{Kovacevic2010}. Several investigations have found that the optical Fe II emission region is stratified. They suggested that Fe II lines arise in two different emission regions, one that produces broad optical Fe II and one that produces narrower Fe II lines \citep{Veron-Cetty2004, Dong2011, Park2022}. On the other hand, through the reverberation mapping technique difficulties have been found in precisely determining the site of Fe II formation in the AGN structure, so different results are obtained in various investigations \citep[see e.g.][]{Kuehn2008, Barth2013, Hu2015, Hu2020}.

The amount of optical Fe II line flux significantly differs in different AGN spectra. It is much stronger relative to Balmer lines in the spectra of narrow line Seyfert 1 galaxies (NLSy1s) than in AGNs with very broad emission lines. The anti-correlations of the Fe II/H$\beta$ ratio with the width of the broad H$\beta$ and equivalent width of the [O III] lines are both part of eigenvector 1 (EV1), which was obtained with the principal component analysis (PCA) done by \cite{Boroson1992}. These anti-correlations are very intriguing since their underlying physics are still not well understood, and they are probably caused by several factors, including the Eddington ratio \citep[see e.g.][]{Marziani2001, Shen2014, Marziani2018, Panda2019}.  Generally, in the case of some AGNs, especially NLSy1s, the emission of Fe II lines is surprisingly strong. It has been estimated that their emission could even be $\sim$ 30\% - 50\% of the total emitted energy from the BLR \citep[see][]{Wills1985, Joly1987, Collin1988}, which is unexpected considering the assumed abundance of iron in BLR emission gas.

Besides being much stronger than expected, the Fe II emission observed in AGN spectra differs significantly in shape from the Fe II spectrum observed for laboratory ionised gas. The difference is in the appearance of forbidden and semi-forbidden Fe II multiplets in AGN spectra \citep{Veron-Cetty2004} and in different relative intensities among some Fe II lines. Namely, in laboratory spectra, obtained in relatively optically thin plasma, the relative intensities among Fe II lines are as expected by theoretical calculations using their atomic parameters. However, in the AGN spectra, the relative intensities among some Fe II lines are violated and do not follow these theoretically expected values. The most striking difference between Fe II emission in laboratory ionised gas for optically thin plasma and that observed in AGN spectra is in the ratio between the UV and optical Fe II emission lines. Namely, in laboratory ionised gas and in theoretical calculations, the total UV Fe II emission is several orders of magnitude larger than optical Fe II emission, while in AGN spectra, the fluxes of the UV and optical Fe II emission lines are of the same order of magnitude \citep{Joly1981}.

To explain the strong Fe II emission in AGN spectra and the excess  flux of optical Fe II lines relative to UV Fe II lines, several mechanisms of excitation have been proposed. It has been  found that resonance fluorescence of continuum radiation as well as photoionisation and recombination are not sufficiently efficient processes to explain strong Fe II emission \citep{Phillips1978a, Netzer1980}. Several studies have come to the conclusion that the standard  photoionisation model of excitation, which is successful at interpreting other broad emission lines, fails to explain strong Fe II emission \citep{Wills1985, Joly1987, Sigut1998, Collin2000}, especially the UV Fe II to optical Fe II ratio \citep{Baldwin2004}. On the other hand, collisional excitation was investigated by \cite{Collin1980} and \cite{Joly1981}, and they found that it can explain observed Fe II emission in the case of the large optical depth of UV Fe II lines and the high column density (N$_e$ > 10$^{23}$ cm$^{-2}$). \cite{Netzer1980} proposed a mixed radiative-collisional excitation mechanism, that is, collisional excitation of Fe II in photoionised gas around AGNs. \cite{Collin2000} gave the model of non-radiative heating due to shocks with an overabundance of iron, and \cite{Sigut1998} proposed and demonstrated the necessity of considering the Ly$\alpha$ ﬂuorescence as an additional mechanism to enhanced the Fe II emission. 

Considering the excess of optical Fe II emission relative to UV Fe II in AGN spectra, the model of large optical depth for UV Fe II lines came closest to the observed values of this ratio. In this model, a large optical depth and a high column density cause a large number of absorptions and reemissions of photons on their optical path, which results in the self-absorption of UV Fe II and an increase in optical Fe II emission \citep{Phillips1978a, Collin1980, Joly1981}. In this model, the large difference of several orders of magnitude between the previously calculated  UV Fe II/optical Fe II ratio and the observed one in the AGN spectra is significantly reduced. However, even in complex models, which include the newest atomic data, the predicted UV Fe II/optical Fe II ratio is still one order of magnitude larger than the observed one \citep{Pandey2024}.

Different theoretical approaches to Fe II emission have resulted in several Fe II models in the literature \citep[see][]{Verner1999, Tsuzuki2006, Bruhweiler2008, Pandey2024, Zhang2024}. However, these calculated models still cannot completely describe the observed relative intensities among Fe II lines in AGNs, especially between UV and optical Fe II lines. On the other hand, in empirical Fe II templates, the relative intensities of the optical Fe II lines are measured using some prototype spectrum with narrow and prominent Fe II lines \citep[see][]{Boroson1992, Veron-Cetty2004, Dong2008, Park2022}, and these relative intensities are fixed when applying the template to different objects. Since the relative intensities of the Fe II multiplets vary for different AGNs \citep{Kovacevic2010,Shapovalova2012,Ilic2023}, it could be expected that these templates will be mismatched with some objects that have significantly different properties for the Fe II lines than in the prototype objects \citep[see Appendix B in][]{Kovacevic2010}.

The semi-empirical multi-component model given in \cite{Kovacevic2010}, supplemented with additional lines in \cite{Shapovalova2012}, separates optical Fe II lines according to the same lower term of transition into five Fe II line groups (P, F, S, G, and H). The relative intensities of the lines within each group were calculated using oscillator strength and temperature \citep[see formula (1) given in][]{Kovacevic2010}, and each line group has a different parameter of intensity, which gives flexibility to the template.
This Fe II model successfully calculated the relative intensities of the majority of Fe II lines in the 4400–5600 \AA \ range ($\sim$ 70 \% of optical Fe II flux), while $\sim$ 30 \% of optical Fe II flux could not be calculated with this approach. The calculated intensities of  these disputed Fe II lines were significantly smaller than the line intensities in the observed spectra. For that reason, \cite{Kovacevic2010} added to the template the line intensities measured in the I Zw 1 object in missing places. This empirically constructed line group is called the 'I Zw 1 Fe II line group'. In this way, they obtained a complete Fe II model with six Fe II line groups: five groups with calculated intensities and a single group with measured line intensities. In \cite{Ilic2023} a number of lines from this group were assumed to be emitted from the high-energy levels (see their Fig. 2). However, the origin of these lines and the atomic processes that are responsible for the emission of their unexpectedly strong flux are open questions. 

In this work, we explore the atomic processes that produce peculiar properties of the Fe II emission in AGN spectra and how they are connected with the macroscopic physical properties of AGN emission regions. We are especially focused on the Fe II lines whose relative intensities could not be explained with the model given in \cite{Kovacevic2010} ('I Zw 1 Fe II line group'), and we try to explain their origin. For the purposes of this research, we modified the template given in \cite{Kovacevic2010} by adding more free parameters to the disputed Fe II line group in order to obtain more empirical clues about the properties of these lines in a large sample of AGNs. In Sect. \ref{Sec2}, we give an overview of the atomic properties and different transitions that occur in the complex Fe II spectrum. In Sect. \ref{Sec3}, we describe our sample selection, method of analysis, and modified Fe II template. The results of the data analysis are presented in Sect. \ref{Sec4} and discussed in Sect. \ref{Sec5}. We outline our conclusions in Sect. \ref{Sec6}.

\section{Multiplets in optical Fe II lines - Atomic characteristics}\label{Sec2}

Understanding the atomic properties of the Fe II multiplets is needed in order to make an appropriate model of the Fe II complex emission in various AGN spectra and to understand which processes might be included in violation of their relative intensities. Therefore, we classified all the Fe II multiplets identified in the 4000–5600 \AA \ range according to the following or by breaking the selection rules. These multiplets could be divided into allowed, intersystem, intercombination, and forbidden \citep{Johansson1984}. The allowed Fe II multiplets are 27 (b$^4$P - z$^4$D$^{o}$), 37 (b$^4$F - z$^4$F$^{o}$), 38 (b$^4$F - z$^4$D$^{o}$), 49 (a$^4$G - z$^4$F$^{o}$) and 42 (a$^6$S - z$^6$P$^{o}$). These allowed transitions are between levels of different parity, following selection rules $\Delta$L = 0, $\pm$1 and $\Delta$S = 0, while intersystem and intercombination multiplets have violations of the LS selection rules \citep{Johansson1984}. The coefficients of spontaneous emission of allowed transitions are of the order of 10$^5$ s$^{-1}$, while for multiplet 42 they are of the order of 10$^6$ s$^{-1}$. 

 Intersystem multiplets have violated the $\Delta$L selection rule, i.e., in the case of multiplets 28 (b$^4$P - z$^4$F$^{o}$), 48 (a$^4$G - z$^4$D$^{o}$) and 32 (a$^4$H - z$^4$F$^{o}$), $\Delta$L = 2, and in the case of the multiplet 41 (a$^6$S - z$^6$F$^{o}$),  $\Delta$L = 3. The coefficients of transition for the multiplets 28, 48, and 41 are similar to those for the allowed lines: $\sim$10$^5$ s$^{-1}$, while for the multiplet 32, it is significantly lower: $\sim$10$^3$ s$^{-1}$. The similar coefficients of spontaneous emission of allowed and intersystem transitions are probably caused by the characteristic of the intersystem multiplets that their levels could be represented as a linear combination of the different state functions. Namely, if two levels have the same parity and the same J value, and if their energies are close, they may mix each other's properties and act as two identical levels \citep{Johansson1984}. Therefore, intersystem transitions are very similar to those allowed in spite of the present violation of selection rules.
 
 Intercombination multiplets have violated the $\Delta$S selection rule. The identified intercombination multiplets are 25] (b$^4$P - z$^6$F$^{o}$), 35] (b$^4$F - z$^6$F$^{o}$), 36] (b$^4$F - z$^6$P$^{o}$), 43] (a$^6$S - z$^4$D$^{o}$), 50] (a$^4$G - z$^4$P$^{o}$), 55] (b$^2$H - z$^4$F$^{o}$) and 56] (b2H - z$^4$D$^{o}$), with $\Delta$S = 1. The coefficients of the spontaneous emission are of the order of 10$^3$-10$^4$ s$^{-1}$. 
Although these transitions have a lower probability than allowed and intersystem transitions, their contribution to AGN spectra could be significant, as in the case of the 43] and 55] multiplets. Multiplet 35] can be strong when the optical thickness is large \citep{Veron-Cetty2004}. Intercombination multiplets 25], 26], 35], and 36] are absent or weak in I Zw 1, while they are relatively strong in IRAS 07598+6508 \citep{Veron-Cetty2006}.

The forbidden lines that were investigated belong to multiplets [4F] (a$^6$D - b$^4$P), [6F] (a$^6$D - b$^4$F), [7F] (a$^6$D - a$^6$S), [19F] (a$^4$F - a$^4$H) and [21F] (a$^4$F - a$^4$G). They all arise in transitions between levels of the same parity, although in certain multiplets some other selection rules are broken as well. The coefficient of the spontaneous emission is in order 1 s$^{-1}$. These lines could be emitted only if electron density is less than 10$^7$ cm$^{-3}$ \citep{Collin1979}, or 10$^8$ cm$^{-3}$ \citep{Veron-Cetty2004}. 
 
The special groups of the Fe II multiplets are those with high energy of excitation, with upper levels having energies of $\sim$ 10 eV. 
These lines could be excited by photon absorption or recombination of the ion Fe III. However, the concentration of the Fe III ion in observed AGNs is small. On the other hand, photon excitation is the most efficient if it is resonant absorption of the strong emission line, in the first place Ly$\alpha$. Detailed analysis of the Fe II lines in AGN spectra given in \cite{Veron-Cetty2004} and \cite{Park2022}, as well as analysis of the excitation of the Fe II high energy levels by Ly$\alpha$ absorption  \citep{Sigut2003,Marinello2016}, indicate that Fe II emission lines that originate from high energy levels, as well as forbidden lines, represent only a small amount of the Fe II flux in optical Fe II lines \citep{Tsuzuki2006}.

\section{Sample selection and method of analysis}\label{Sec3}

To investigate the properties of the Fe II emission lines, we obtained the sample of AGN spectra from the Sloan Digital Sky Survey (SDSS)\footnote{\url{https://www.sdss4.org/dr16/}} database, Data Release 16 (DR16)  \citep[see][]{Ahumada2020}. DR16 was released in 2019, and it is the fourth data release of the fourth phase of the Sloan Digital Sky Survey (SDSS-IV). It contains the optical single-fibre spectroscopy observations over millions of QSO spectra, made with the 2.5 m Sloan Foundation Telescope at the Apache Point Observatory. 

Using the Structural Query Language (SQL), we searched the SDSS DR16 database and chose the spectra of the AGNs that satisfied the following criteria:

\begin{enumerate}[(i)]

\item   Objects are to be Type 1 AGNs  ('QSO' spectral class in SDSS spectral classification).

\item   The cosmological redshift (z) is to be z $<$ 0.7 in  order to cover optical Fe II emission in the 4000-5600 \AA \ range.

\item   The median S/N per pixel of the whole spectrum is greater than 30, and the median S/N per pixel in the g-band is greater than 30 in order to get high-quality spectra in which Fe II lines can be precisely fitted.

\item   Objects have no problems with redshift determination (z warning = 0).

\end{enumerate}

In this manner, we obtained 1291 AGN spectra. Afterwards, we used visual inspection to reject all spectra with strong absorption lines, which significantly affect emission lines in the 4000–5600 \AA \ spectral region. Finally, our sample contains 1046 high-quality spectra of Type 1 AGNs , which are convenient for sophisticated analysis of the complex Fe II emission in the 4000-5600 \AA \ region.

\begin{figure}
        \centering
        \includegraphics[width=85mm]{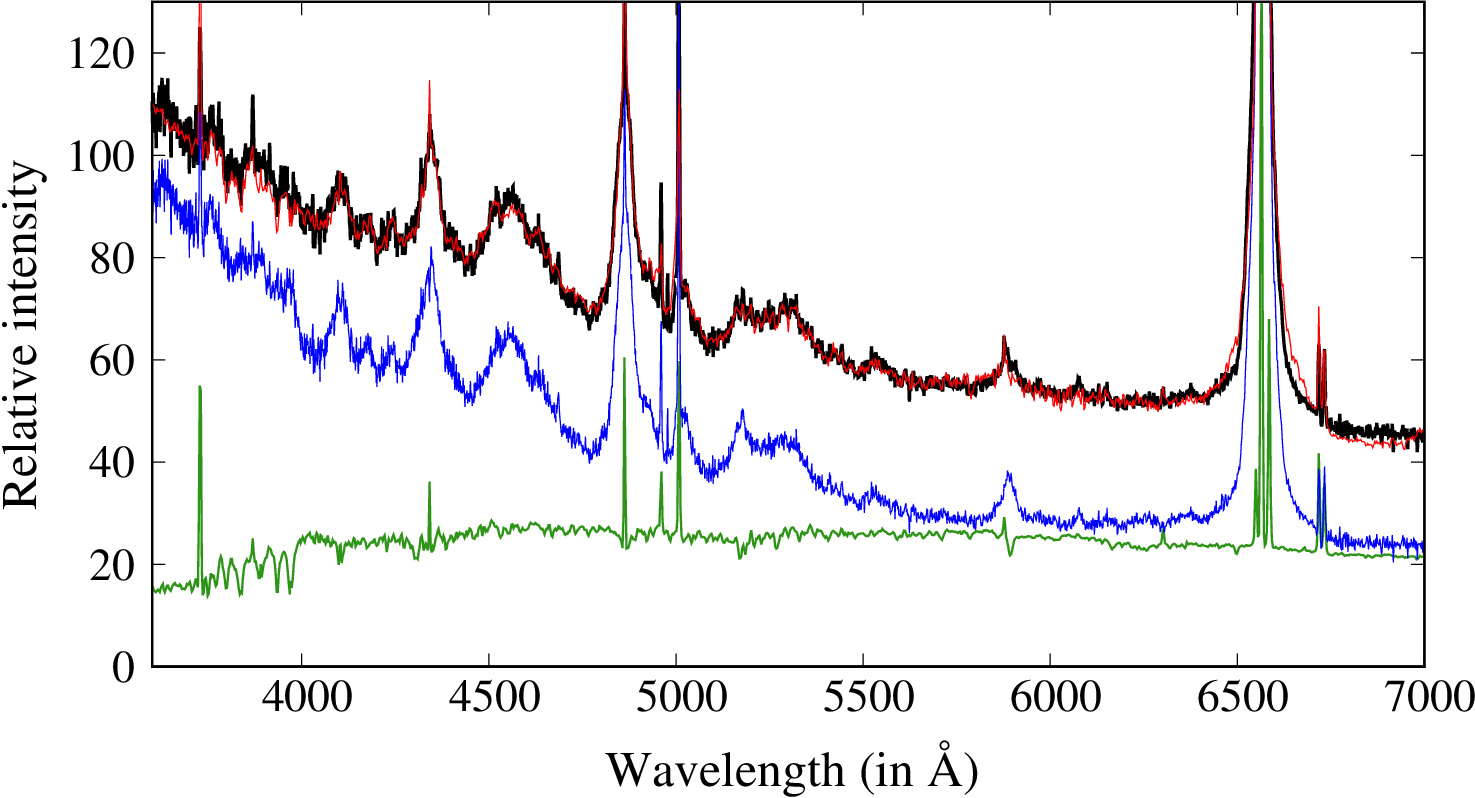}
    \caption{Example of the decomposition of the spectrum (SDSS J105355.69+661202.2) to the host galaxy and the pure AGN contribution using spectral PCA. The black line is the original spectrum, and the red line is the best fit obtained with a linear combination of 15 eigenspectra. The green line is the reconstructed host galaxy contribution, and the blue line is the pure AGN contribution (host contribution subtracted from the observed spectrum). }
    \label{fig01}
\end{figure}
\begin{figure}
        
        \includegraphics[width=65mm]{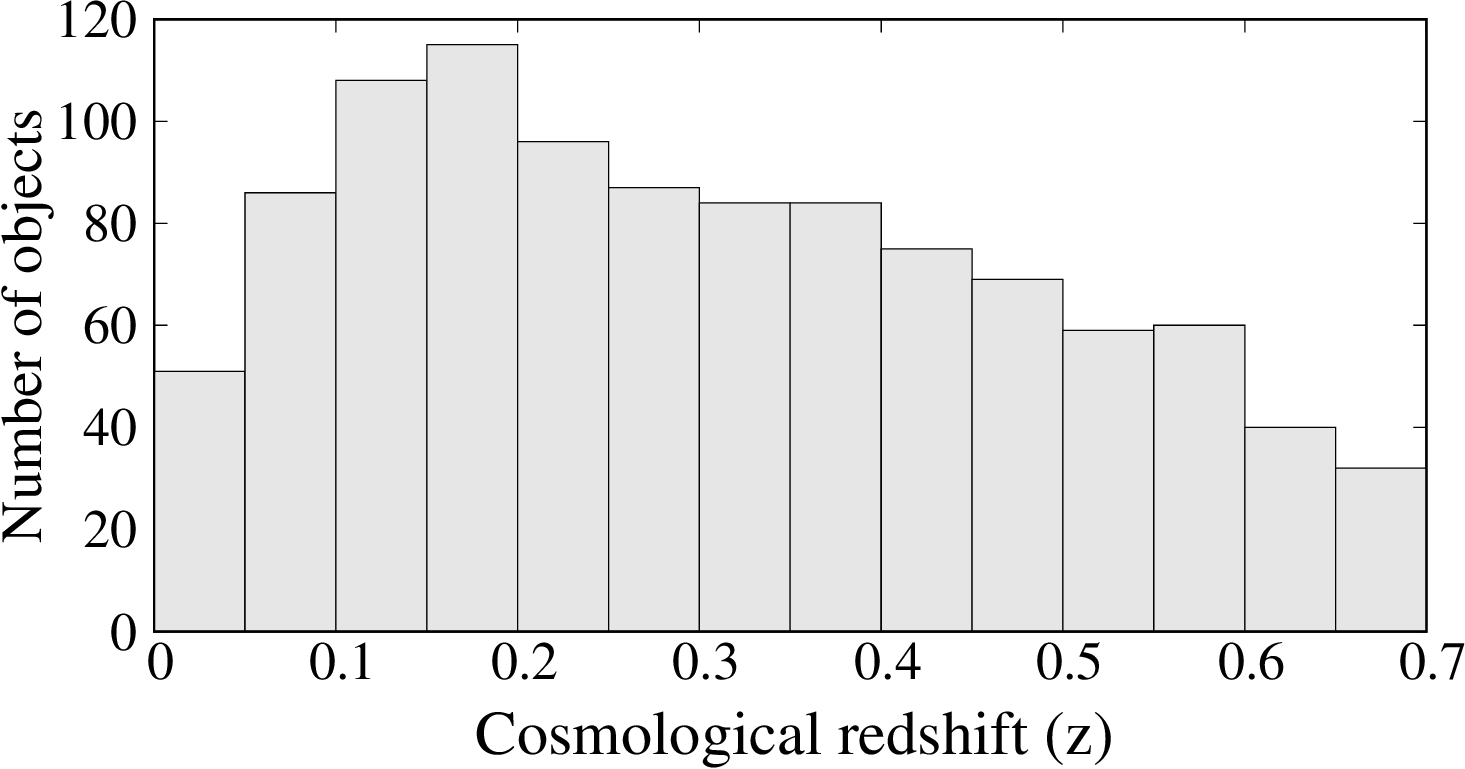}
        \centering
    \caption{Histogram of the cosmological redshift (z) distribution in a sample of 1046 AGNs. }
    \label{fig02}
\end{figure}

To correct Galactic reddening, we used the standard extinction law given in \cite{Howarth1983} and extinction coefficients given in \cite{Schlegel1998}, available from the NASA/IPAC Infrared Science Archive (IRSA).\footnote{\url{http://irsa.ipac.caltech.edu/applications/DUST/}} We corrected  the spectra for the cosmological redshift, and we applied spectral PCA in order to decompose the spectra into the host galaxy and the AGN contribution,  as it is done in various investigations of a large SDSS AGN sample \citep[see e.g.][etc.]{Vanden2006, Shen2015, Shen2019, Rakshit2020, Ren2024}. We followed the procedure described in Vanden Berk et al. (2006), who found that diverse AGN spectra could be well decomposed into pure QSO and host galaxy contribution using two independent sets of eigenspectra, derived from the pure-QSO \citep{yip04a} and the pure-galaxy \citep{yip04b} SDSS samples. In this method, the observed spectra are fitted with a linear combination of  pure-galaxy and pure-QSO eigenspectra together, and the result of the best fit are the coefficients that multiply the eigenspectra in order to match the observations.  The host galaxy contribution in spectrum is then estimated as the linear combination of only pure-galaxy eigenspectra, obtained from the best fit. \cite{Vanden2006} found that this technique can well reproduce the host galaxy spectrum, its contributing fraction, and its classification in large sample of various AGNs.

We used the first ten QSO eigenspectra and the first five galaxy eigenspectra given in \cite{yip04a, yip04b} and performed the non-parametric fitting of a linear combination of 15 eigenspectra on our sample \citep[for more details see][]{Lakicevic2017}. The host-galaxy spectra are obtained as the linear combination of the 5 galaxy eigenspectra with linear coefficients obtained from the best fit. The pure AGN spectra were obtained when host-galaxy spectra, with previously masked narrow emission lines, were subtracted from the observed spectra. Since our sample is chosen to have a large signal-to-noise ratio and no strong absorbtion lines, most of the spectra have weak galaxy contributions. The example of the decomposition of the AGN spectrum to host and the pure AGN contribution is given in Fig. \ref{fig01}. The distribution of the cosmological redshift of the sample is given in Fig. \ref{fig02}. 

After host galaxy subtraction, the pure AGN spectra are fitted in 4000-5600 \AA \ range, applying the multi-Gaussian decomposition and using the $\chi^2$ minimalisation routine to obtain the best-fit parameters, similarly as done in \cite{Popovic2004, Kovacevic2010, Kovacevic2015}. The Balmer lines are fitted with three Gaussians. The narrow Gaussian represents emission from the narrow line region, while the broad Balmer lines are fitted with two-component model. In that model, one Gaussian fits the core of the broad Balmer line and represents the emission from the ILR, while the other Gaussian fits the wings of the line and represents the emission from the very broad line region (VBLR), which is assumed to be the inner layer of the BLR. This decomposition was applied to all spectra from the sample, regardless of whether they belong to  NLSy1 or Broad Line Seyfert 1 objects. There are several studies which found that the Lorentzian can fit the profiles of the broad Balmer lines in NLSy1 galaxies very well \citep[see e.g.][]{Cracco2016,Marinello2016}. However, we found that the broad Balmer lines in these objects could be also very well fitted with sum of two Gaussians, where the narrower ILR Gaussian is of significantly higher intensity than the broader VBLR Gaussian, which fits the wings. In this way, we were able to apply the same decomposition model to the entire sample.

The difference compared to the fitting decomposition given in \cite{Kovacevic2015} is that we reduced the number of fitting parameters for Balmer lines (H$\beta$, H$\gamma$, and H$\delta$) in order to more precisely fit the continuum level and the Fe II lines, which overlap with Balmer lines. Also, we included in the fitting model the He I 4027.3 \AA \ and [S II] 4069.7, 4077.5 \AA \ lines (for more details on fitting decomposition, see Appendix \ref{A}). Similarly, as in previous investigations, all narrow lines (narrow components of Balmer lines, [S II] and [O III]) are fitted  with the same parameters of shift and width, while in the case of asymmetry in [O III]4959, 5007 \AA, they are fitted with two Gaussians. 

The main difference in the decomposition model compared to the ones given in \cite{Kovacevic2010} and \cite{Kovacevic2015}, is applying the new, complex Fe II template, adapted to the needs of this research, where more free parameters are given for the Fe II lines whose observed relative intensities do not follow theoretically expected values.

\subsection{Modified optical Fe II template}\label{3.1}

Since the aim of this investigation is to understand the processes that produce peculiar properties of the Fe II emission in AGN spectra, we divided all optical Fe II lines in the  4000-5600 \AA \ range into two large groups. 
The first group are optical Fe II lines, whose observed relative intensities are not as expected by theoretical calculations following their atomic parameters. These lines are referred to further in the text as inconsistent Fe II lines (Fe II$_{incons}$), which alludes to the inconsistency of their measured intensities in spectra with theoretically expected values. This group includes the Fe II lines whose intensities could not be calculated in the model given in \cite{Kovacevic2010}, so they were measured empirically ('I Zw 1 Fe II line group'), but there are also some additional lines whose spectra are stronger than expected according to their transition probabilities. On the other hand, the Fe II lines whose relative intensities calculated following their atomic parameters are consistent with observations are referred to as consistent lines (Fe II$_{cons}$).

For the purpose of this research, we modified the template given in \cite{Kovacevic2010} and supplemented it in \cite{Shapovalova2012} in order to use it as a sophisticated tool for the investigation of the unexplained optical Fe II lines. The modification is done in two directions: 
\begin{enumerate}[(i)]

\item Improving the template by addressing deficiencies in the initial model given in \cite{Kovacevic2010}. We focused on two problems. The first is the determination of the Fe II pseudocontinuum \citep[see][]{Popovic2019}, which is improved by including a double-Gaussian model for fitting the strongest Fe II lines; the second is correcting the intensities of empirically measured inconsistent Fe II lines, mostly near the H$\gamma$ line. 

\item  Adding the new parameters of freedom in the Fe II template for inconsistent Fe II lines, which is necessary in order to better understand the underlying physical processes that are responsible for their emission.
 
\end{enumerate}

The majority of Fe II lines that were initially part of the F, S, and G groups in \cite{Kovacevic2010} belong to the consistent Fe II lines. In this work, they were modelled as in \cite{Kovacevic2010}: the relative intensities of lines within each group are calculated, and each group has a free parameter of intensity. While in previous investigations they were fitted with single-Guassian for each line, in this investigation these lines were fitted with double-Gussians, similarly to Balmer lines, where the ILR component fits the core and the VBLR component fits the wings of the lines. All consistent lines have the same width of ILR components, and the same width of VBLR components, and these values are not tied with Balmer line ILR and VBLR components. Also, the intensity ratio of ILR and VBLR Fe II components is the same for all consistent lines.

The rest of the identified lines ('I Zw 1 Fe II' line group, P and H groups) are assigned as inconsistent lines. Namely, in \cite{Kovacevic2010} and also in \cite{Ilic2023}, it was supposed that 'I Zw 1 group of lines', which should be negligible according to their transition probabilities, appears since the lines with very similar wavelengths arise from high energy levels and overlap with these weak lines from lower levels, in this way making the stronger Fe II emission. In this work, we adopted a different strategy without involving lines with high energy levels, or at least without giving crucial roles to these lines. We found that the inconsistent lines could be identified as lines whose energies at the upper levels do not exceed 6 eV and which belong to F or G groups, with the addition of only a few semi-forbidden, forbidden, or lines that originate from higher levels of excitation. Therefore, we suppose that the majority of disputed lines are actually lines that arise from the same levels as consistent lines, just amplified with some unknown atomic processes. We noticed that some of the Fe II lines from the P group given in \cite{Shapovalova2012} do not follow relative intensities calculated with formula (1) in \cite{Kovacevic2010} (mostly in H$\gamma$ region), and we joined these lines to the inconsistent lines as well. Also, the lines from group H given in \cite{Shapovalova2012} are joined to inconsistent lines since they belong to semi-forbidden transitions, and according to their atomic parameters, they should have negligible intensity relative to consistent lines, but they could be well seen in many spectra.

To investigate in more detail the inconsistent lines, we divided them into several line groups according to their atomic properties and specific behaviour observed in numerous spectra of the sample. The relative intensities of the lines within groups are fixed, as measured in several NLSy1 spectra from this sample (see Appendix \ref{B}), while the intensities of the inconsistent line groups are free parameters. The inconsistent Fe II lines are fitted with a single-Gaussian model, and each inconsistent line group has a free parameter for the width in order to investigate the site of their origin. The shift of inconsistent lines is assumed to be the same as that of consistent lines. 

The inconsistent Fe II groups were formed as follows:

\begin{enumerate}

\item The P+ group is formed of lines that arise in allowed transitions (multiplet 27) and intersystem transitions (multiplet 28) and belong to the P group, with the addition of the three allowed lines whose lower term of transition is ${}^4$F ($\lambda$4620 \AA, $\lambda$4629 \AA \ and $\lambda$4666 \AA). For these three lines, we found that they have similar behaviour as P lines, i.e., their intensities increase/decrease relative to consistent lines similarly to the intensities of the P lines.

\item  The G+ group contains lines that mostly arise in intersystem transitions (multiplet 48) and have a lower term of transition ${}^4$G,  with addition of intersystem multiplet 32, which has similar behaviour.

\item  The H group consists of two semi-forbidden lines (multiplets 56] and 55]). 
 
\item  The other lines (OL) group consists of the lines with the least probability of emission according to the atomic parameters. It consists of two allowed lines with a lower term of ${}^4$F, 12 semi-forbidden lines, 9 forbidden lines, and 4 lines that arise from high energy levels. 
\end{enumerate}

Detailed explanation about the modelling of the consistent Fe II lines with a double-Gaussian model and of the inconsistent Fe II lines with a single-Gaussian component is given in Appendix \ref{B}.
The complete list of the inconsistent lines in the 4000 - 5600 \AA \ range with their empirically measured intensities, as well as the list of the consistent Fe II lines from the F, S, and G groups and their calculated relative intensities, are given in Table \ref{T1}. The simplified Grotrian diagram of inconsistent Fe II groups is shown in Fig. \ref{figA1}.

\begin{figure} 
 \includegraphics[width=87mm]{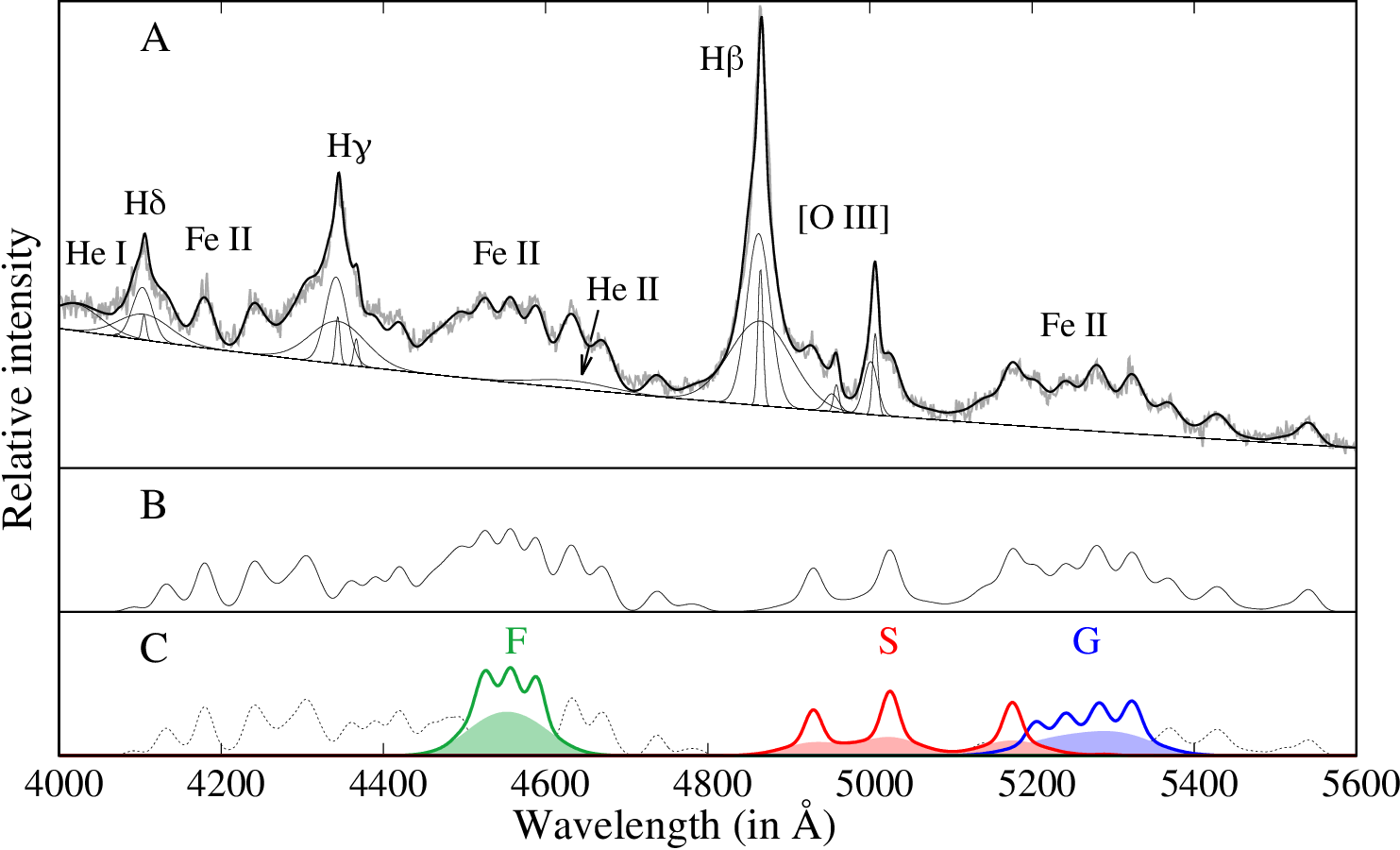}
 \centering
    \caption{Example of the fit in the 4000-5600 \AA \ range for SDSS J111941.12$+$595108.7. Panel A shows the continuum fitted with a power-law, emission lines with a single or multi-Gaussian model, and Fe II lines with a complex Fe II template, which is shown separately in panel B. Panel C shows the inconsistent Fe II lines, denoted with dotted lines, and the consistent Fe II lines, which are coloured in green (F group), red (S group), and blue (G group). The sum of the VBLR components of consistent the Fe II lines is shaded with the appropriate colour for each group.}
    \label{fig02_1}
\end{figure}

The final Fe II model is complex, with 15 free parameters of fit, which enable large flexibility during the fitting following the `breathing' of Fe II lines in various spectra. The intensities of consistent lines (F, S, and G groups) are described with 5 parameters: temperature for calculation of the relative intensities of lines within the groups, 3 parameters of intensity of ILR components for each group, and one parameter for intensity ratio of VBLR/ILR components, which is the same for all these groups. The intensities of inconsistent lines are defined with 4 free parameters for P+, G+, H, and OL groups.   In each of Fe II groups there are at least some lines which do not overlap with the lines from the other line groups, which allows accurate determination of group intensity. The shift is taken to be the same for all lines, while several free parameters for widths are defined: the widths of the ILR and VBLR components of consistent lines, and 4 free parameters of widths for the P+, G+, H, and OL groups.  In the case of the spectra with narrow Fe II lines, the shift and the widths of Fe II lines from different line groups could be fitted with high reliability, since et least some of the Fe II lines from each line group could be well resolved. However, in spectra with very broad (FWHM > 8000 km/s) and weak Fe II lines, where single Fe II lines cannot be resolved, we had to give some additional fitting constrains in order to skip non-physical solutions. We limited Fe II shift to be between -3000 km/s and 3000 km/s, and FWHM of Fe II lines to be less than 15000 km/s. These spectra account for less than 10\% of the sample. In Fig. \ref{fig02_1} we show an example of the fit of an AGN spectrum with a complex Fe II template. The inconsistent groups of the lines are shown separately in Fig. \ref{fig02_2}, panel A.
 The properties of the complex Fe II template are summarised in Table \ref{T2}.

\subsection{Fitting of Fe II lines in the UV 2650\AA \ - 3050\AA \ range}\label{3.2}

To compare the properties of the optical and UV Fe II lines, we also fitted and analysed the UV spectra in the 2650\AA \ - 3050 \AA \ range for the subsample for which the UV range was available due to redshift. After rejecting the spectra with strong absorption in the Mg II 2800 \AA \  line, the subsample with UV spectra contains 338 objects.  These spectra were fitted with the complex UV Fe II model described in detail in \cite{Popovic2019}. This model consists of six Fe II line groups, which have different parameters of intensity, thus enabling flexibility of the model. Five of the six line groups are different multiplets of UV Fe II:  60, 61, 62, 63 and 78. Additional group of UV Fe II lines is the one which relative intensities are taken from I Zw 1 empirically (`I Zw 1 group of lines'). In model given in  \cite{Popovic2019}, all Fe II lines are described with single Gaussian function, and all Fe II lines have the same shift and width. However, for this research, similarly as it is done for optical Fe II lines, we added an additional broad Gaussian to each Fe II line, since we found that two-component approach gives better fit of UV Fe II lines in some spectra. In this way, each of the UV Fe II lines  is described with two Gaussians: ILR and VBLR components. The widths of ILR and VBLR components are the same for all the lines in the template, and they are not tied with the same parameters of optical Fe II. Finally, our UV Fe II model consists of 10 free parameters: six parameters of intensity for each line group, the widths of ILR and VBLR components, the shift, and the parameter which represents the intensity ratio of VBLR/ILR components, which is the same for all lines. With this model, we were able to fit well even the most complex UV Fe II features in this range. Mg II lines were fitted with two Gaussians (one for the wings and one for the core of the line), and Al II line with single Gaussian. The example of fit in UV range is shown in Fig. \ref{figR}.

As it can be seen in Fig. \ref{figR}, the UV Fe II lines in multiplets 62, 63 and 61, overlap with Mg II line, but also with the 
UV Fe II lines from the `I Zw 1 group', which are not well identified and explained. On the other hand, the UV Fe II bump at $\sim$ 2950 \AA, is mostly emission of the multiplet 60, without overlapping with Mg II, and without influence of unexplained UV Fe II lines. Also, it has well defined shape, which improves the confidence of the fit. In some spectra, the multiplet 78 gives the small contribution to the flux of this bump near $\sim$ 2980 \AA, which is less than 10\%. In this work, we used only the sum of multiplets 60 and 78, i.e. the flux of Fe II bump at $\sim$ 2950 \AA \ (see the red line in Fig. \ref{figR}), for comparison with optical Fe II. Since the multiplet 78 is negligible in most of the spectra, and its flux do not exceed 10\% of bump flux in spectra where it is present, the UV Fe II flux through this work is assigned as UV Fe II$_{60}$. 

We checked how representative of the total UV FeII emission is UV Fe II$_{60}$. We found that although there are small variations between the intensities of line groups in different spectra, the flux of total UV Fe II emission in 2650\AA \ - 3050 \AA \ range is in very high correlation with UV Fe II$_{60}$, with Spearman coefficient of correlation r=0.94, and P-value=0.

\subsection{Measurements of spectral properties}\label{3.3}

The Full Width at Half Maximum (FWHM) of the broad H$\beta$ line is measured as the sum of two broad components (ILR and VBLR), obtained from the best fit \citep[see][]{Kovacevic2015}. The FWHM of consistent Fe II lines (FWHM Fe II$_{cons}$) is measured in the same manner: we singled out one Fe II consistent line, and we measured the width at half maximum of its total shape, which is the sum of the ILR and VBLR components. Since the inconsistent lines are fitted with single Gaussians, the widths of lines in different inconsistent line groups are directly obtained from the best fit.  The same procedure applied for measuring the width of the consistent Fe II lines, is done for measuring the FWHM of the UV Fe II lines. We single out one UV Fe II line, and we measured FWHM of the total profile, which includes ILR and VBLR components.

The flux of the optical continuum is determined by measuring the intensity of the power law obtained from the best fit at 5100 \AA, and luminosity is calculated using the formula given in \cite{Peebles1993}, with adopted cosmological parameters: $\Omega_M$=0.3, $\Omega_\Lambda$=0.7 and $\Omega_k$=0, and Hubble constant $\rm H_{0}$=70 km s$^{-1}$ Mpc$^{-1}$. The Eddington ratio (R$_{Edd}$) is estimated as R$_{Edd}$=L$_{bol}$/L$_{Edd}$, where L$_{bol}$ is bolometric luminosity calculated as: $L_{bol} = 9\cdot\lambda L_{5100}$ 
\citep{Kaspi2000}, and L$_{Edd}$ is Eddington luminosity, calculated as L$_{Edd}$=1.26$\cdot$10$^{38}$(M$_{BH}$/M$_{sun}$) erg s$^{-1}$ \citep{Wu2004, Bian2007}.

\begin{figure} 
\centering
 \includegraphics[width=85mm]{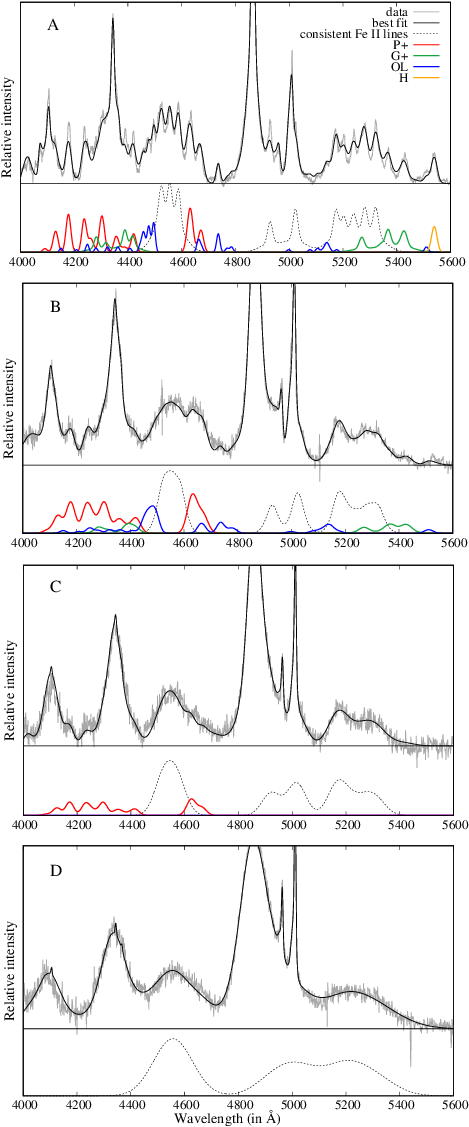}
    \caption{Inconsistent Fe II lines in AGN spectra. In the spectrum shown in panel A, all inconsistent Fe II lines are present and assigned to the P+, G+, OL, and H groups of lines. In the spectrum shown in panel B, the H group of inconsistent lines is missing, while in panel C, only the P+ group of inconsistent lines is present. The spectrum shown in panel D has no inconsistent lines. All spectra are shown after the subtracted power-law continuum. The spectra shown in panels belong to objects SDSS J154732.17+102451.2 (A), SDSS J085632.40+504114.0 (B), SDSS J012302.02-024400.3 (C) and SDSS J002332.33-011444.1 (D).}
    \label{fig02_2}
\end{figure}

\begin{figure}
        
        \includegraphics[width=75mm]{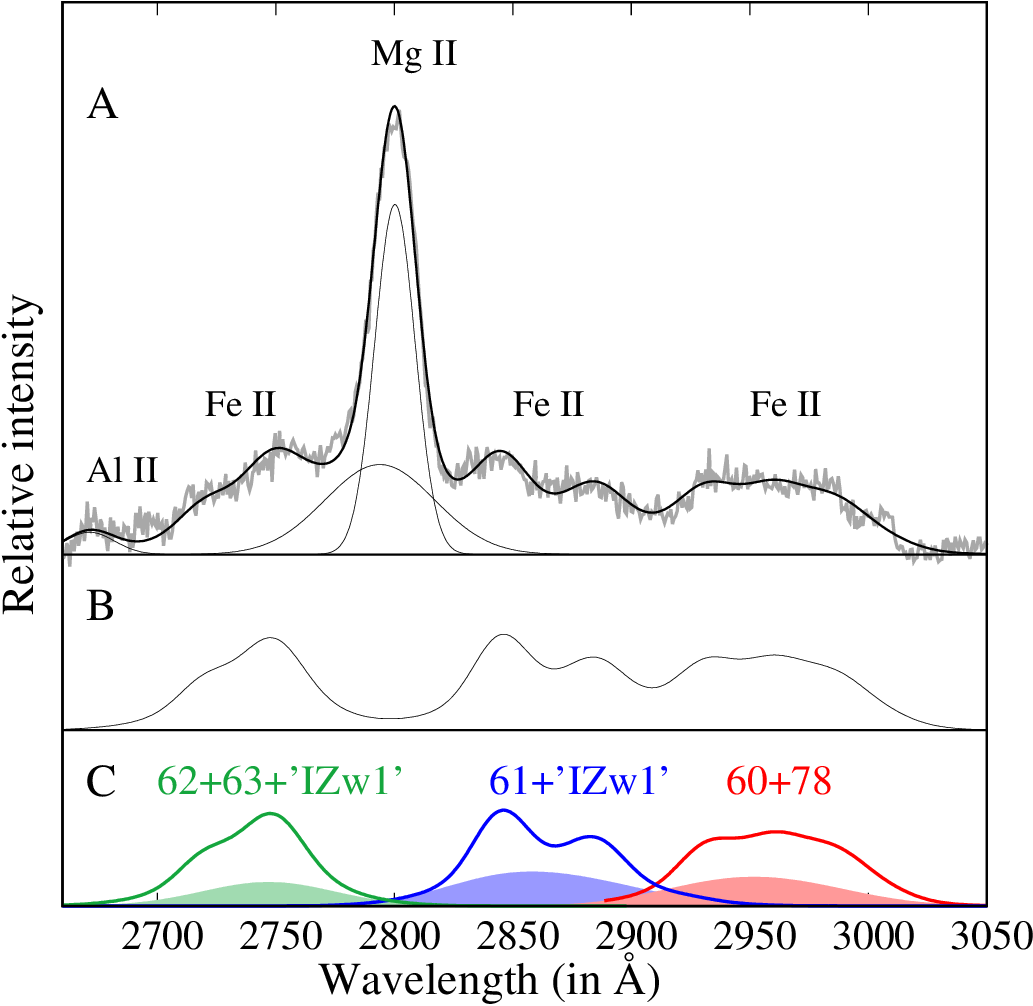}
        \centering
    \caption{ Example of the fit in the 2650-3050 \AA \ range for SDSS J081025.94+410000.9 (plate-mjd-fiber: 8292-57373-0141). Panel A shows the Mg II line fitted with two Gaussians, Al II with a single Gaussian, and Fe II lines with a complex UV Fe II template, which is shown separately in panel B. Panel C shows the UV Fe II lines, which are marked with different colours for the sums of different line groups. The multiplets 62 and 63 and the additional 'I Zw 1 lines' in that range are coloured in green, the multiplet 61  and 'I Zw 1 lines' near $\sim$ 2850 \AA \ are coloured in blue, and the multiplets 60 and 78 are coloured in red. The sums of the VBLR components of UV Fe II lines from different multiplets are shaded with the appropriate colour for each group of lines. }
    \label{figR}
\end{figure}

\begin{figure} 
\centering
 \includegraphics[width=89mm]{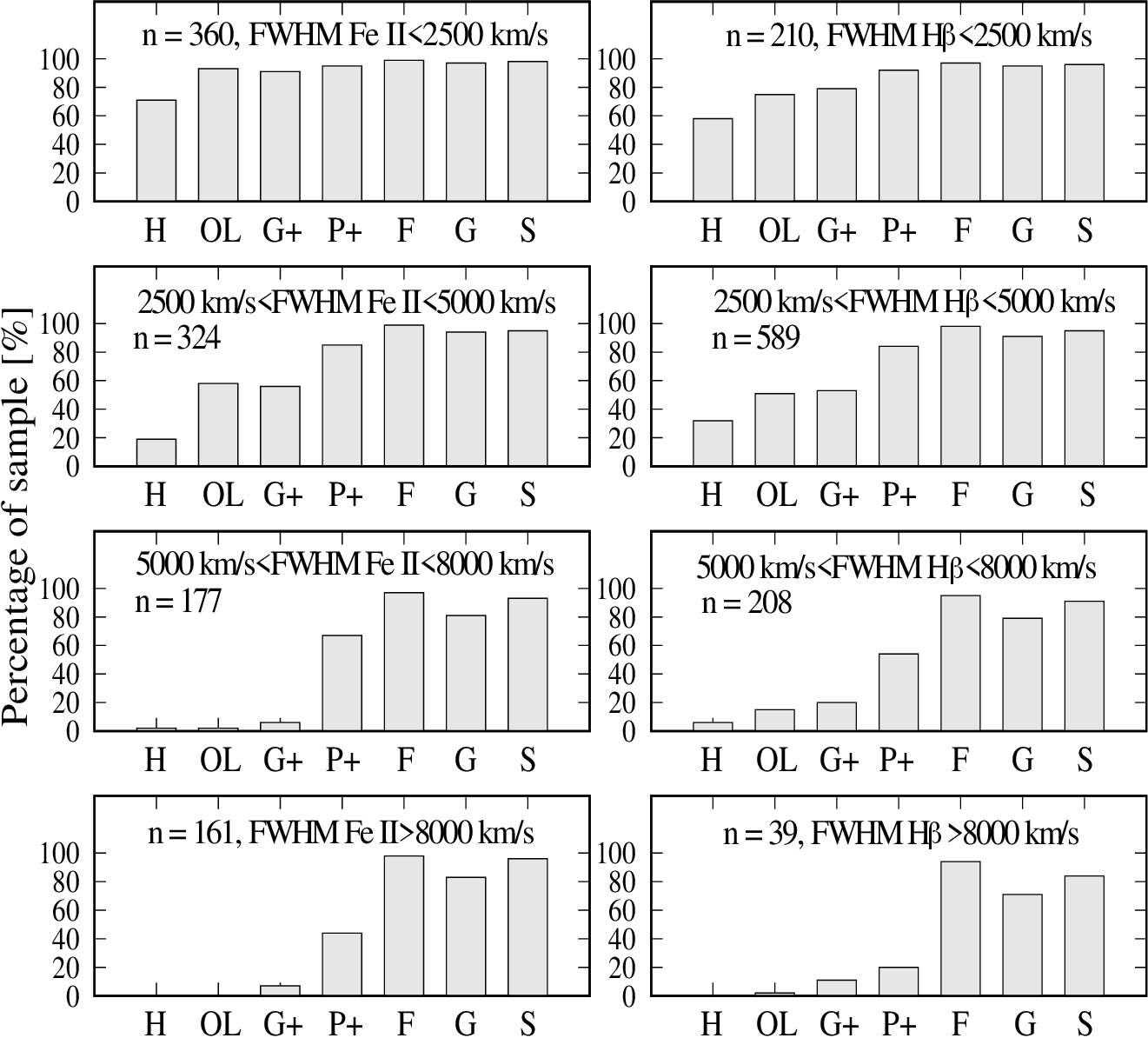}
    \caption{ Presence of the Fe II line groups in different subsets. The sample is divided into several subsets with $n$ spectra following different ranges of line widths (FWHM Fe II or FWHM H$\beta$). The histograms show the percentage of spectra with present Fe II lines from different Fe II groups (H, OL, G+, P+, F, S, G) in each subset.}
    \label{fig03}
    \end{figure}

\begin{figure} 
\centering
 \includegraphics[width=85mm]{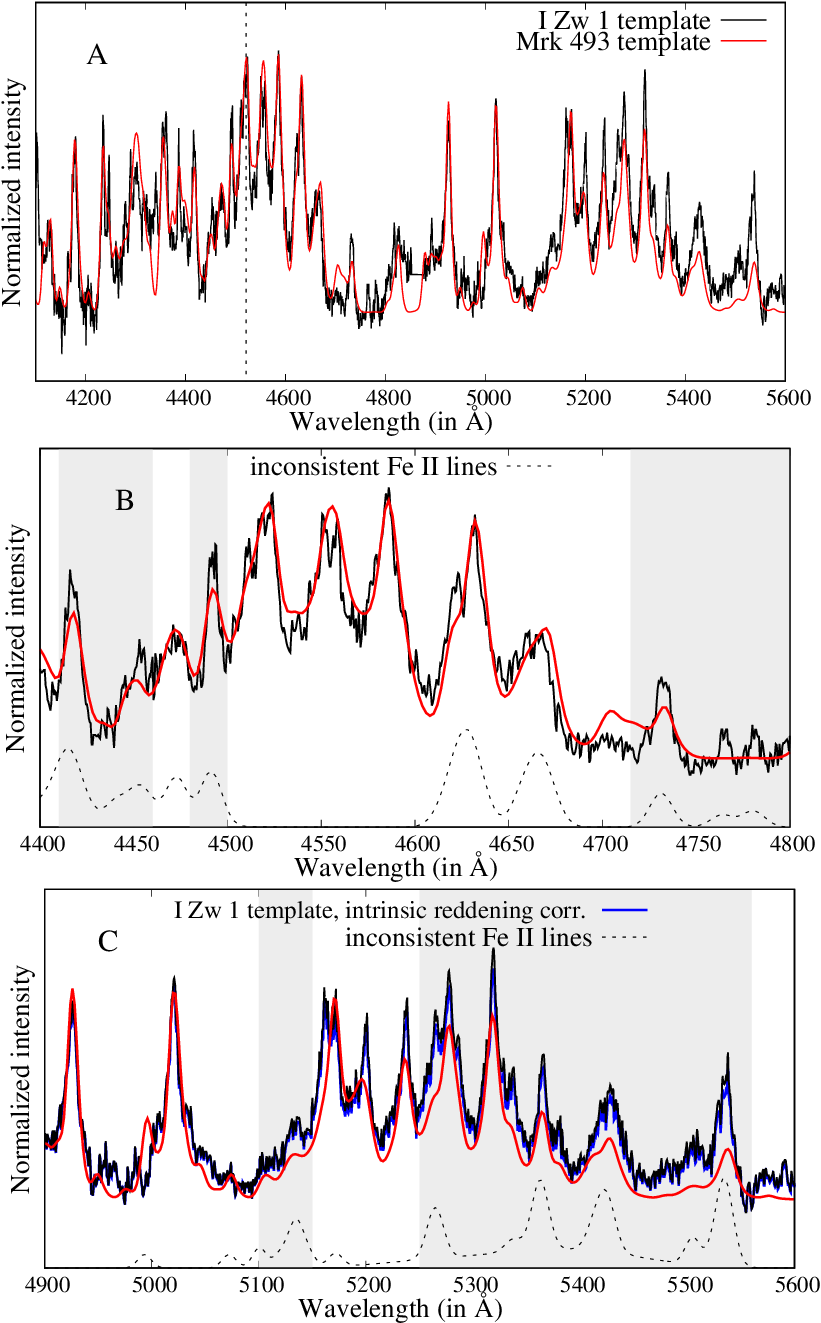}
    \caption{ Comparison of relative intensities of Fe II lines in I Zw 1 and Mrk 493. The black line represents the I Zw 1 Fe II template given in \cite{Boroson1992} and the red line is the Mrk 493 Fe II template given in \cite{Park2022}. Both templates are scaled to have the same intensity of $\lambda$4520 \AA \ line (see vertical dashed line in panel A). The Fe II emission bluewards  to H$\beta$ is shown separately in panel B, and the redwards the H$\beta$ is shown separately in panel C. The shaded regions in panels B and C highlight the mismatch between two templates. The inconsistent Fe II lines obtained from our model for I Zw 1 spectrum are shown with dashed line in bottom of panels B and C. Additionally, the I Zw 1 template corrected for intrinsic reddening is shown in panel C, denoted with the blue line.}
    \label{fig03_1}
    \end{figure}

\begin{figure} 
\centering
 \includegraphics[width=75mm]{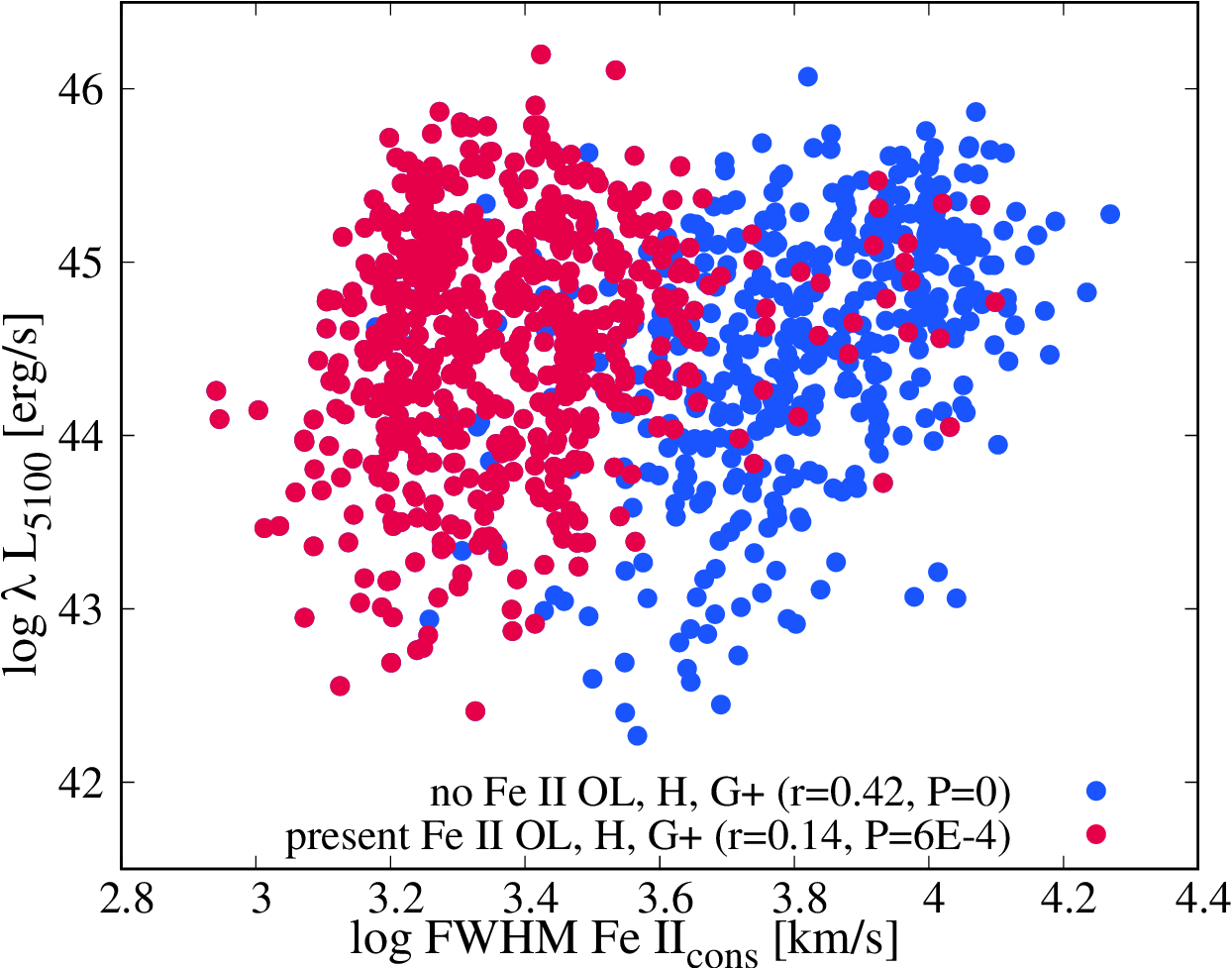}
    \caption{Presence of inconsistent line groups in the sample. The blue dots represent the objects with no G+, OL, and H inconsistent line groups, while the red dots are objects with at least one of these line groups present. The Spearman coefficient of correlation (r) and P-value for log$\lambda L_{5100}$ versus logFWHM Fe II$_{cons}$ are given for each subsample.}
    \label{fig04}
\end{figure}

\begin{figure} 
\centering
  \includegraphics[width=75mm]{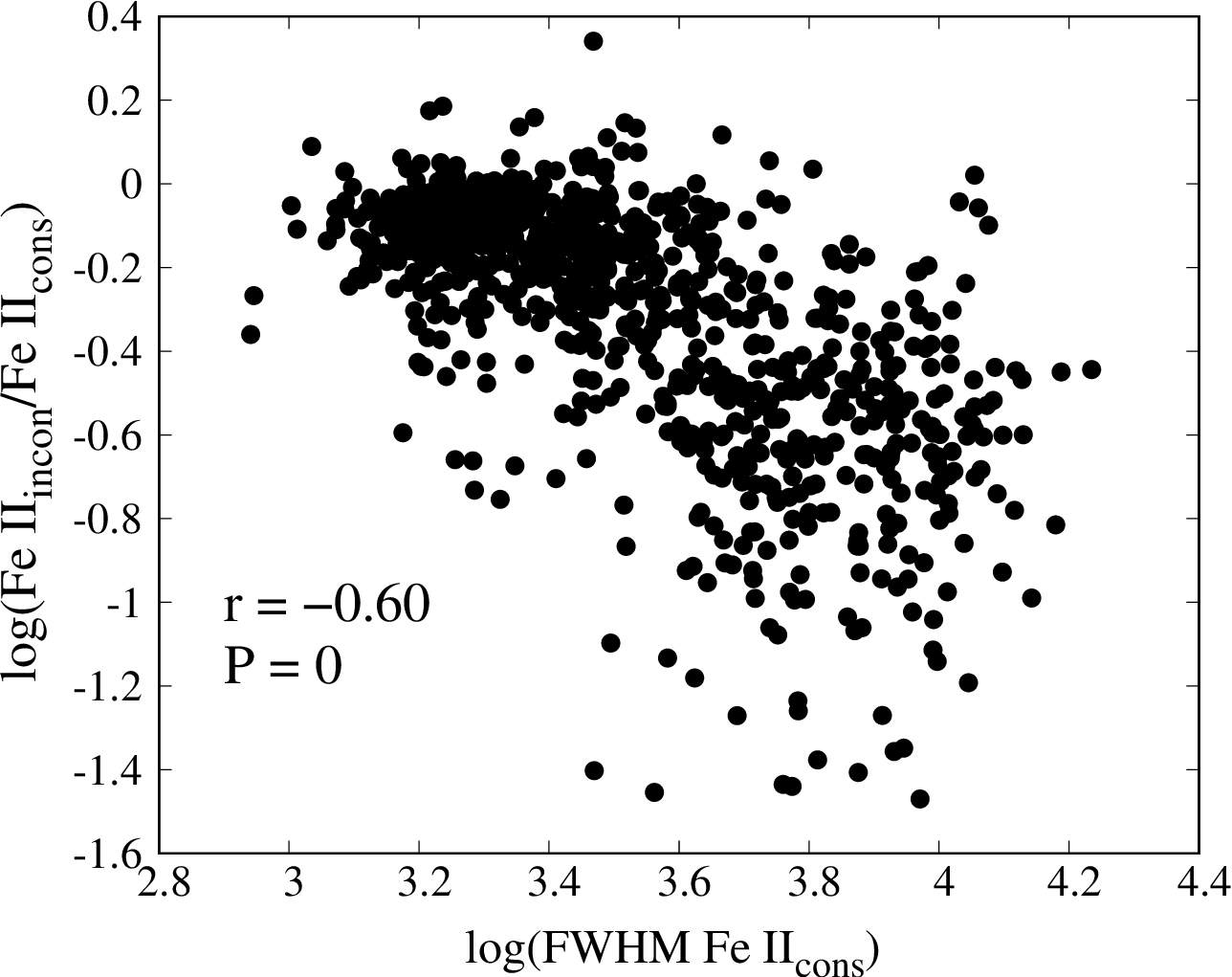}
 \includegraphics[width=75mm]{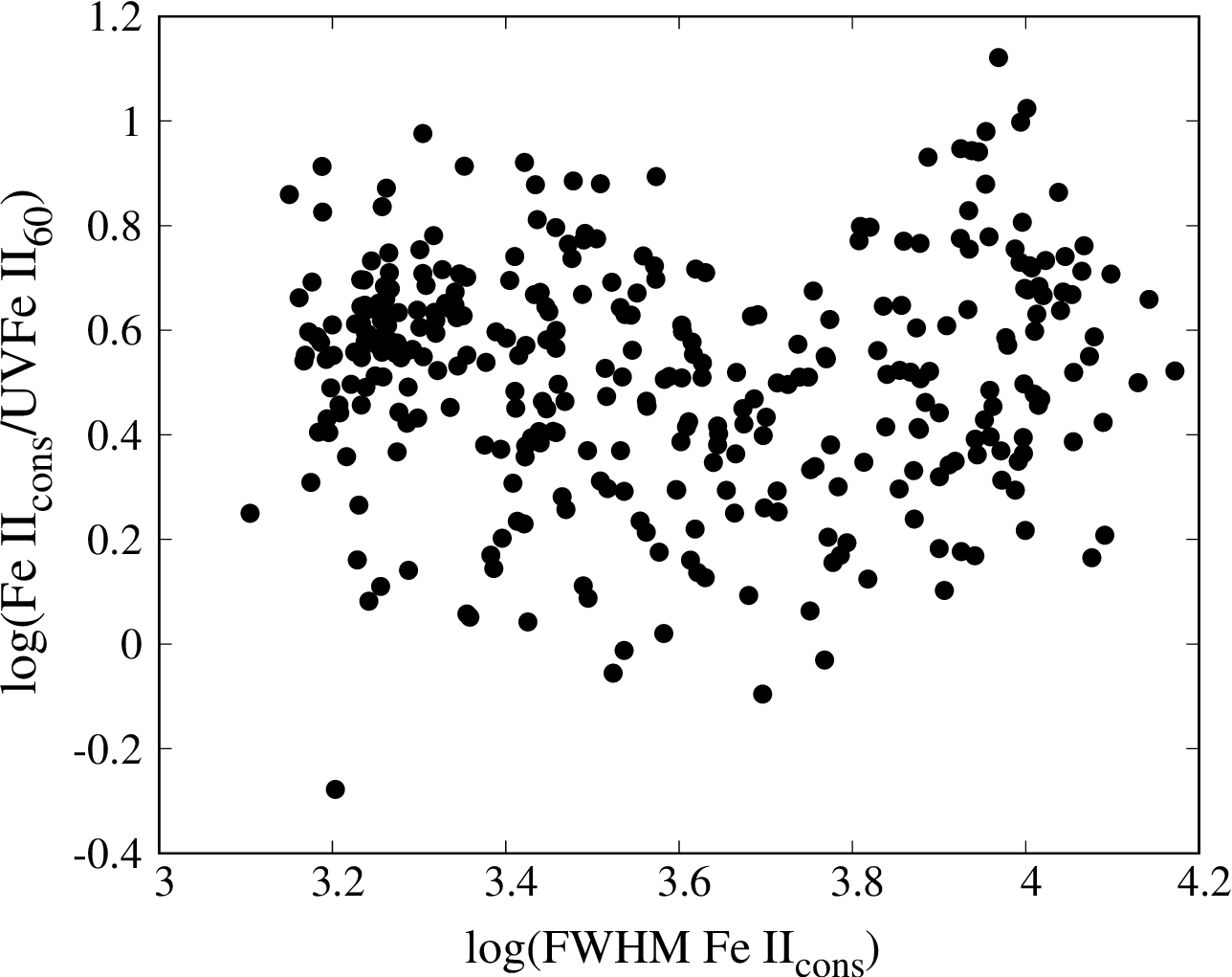}
   \caption{ Correlation between the width of the Fe II$_{cons}$ and Fe II$_{incons}$/Fe II$_{cons}$ ratio (up) and the Fe II$_{cons}$/UV Fe II$_{60}$ ratio (bottom). The width is given in kilometres per second.}
    \label{fig05}
\end{figure}

\begin{figure} 
\centering
 \includegraphics[width=75mm]{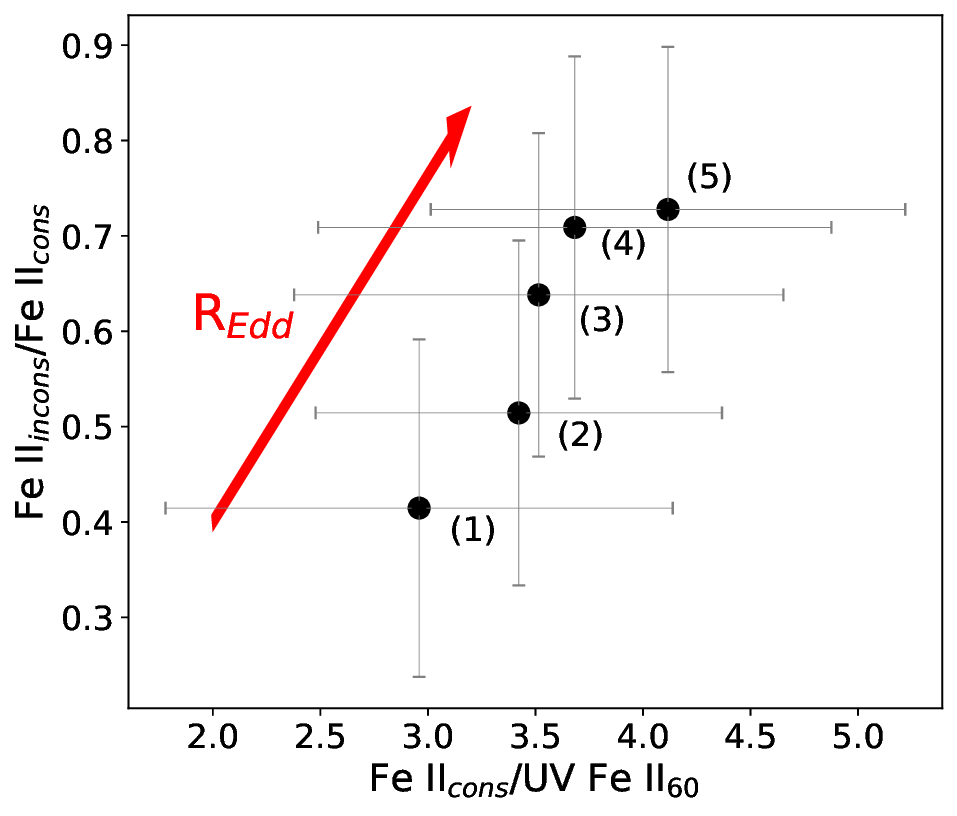}
 \caption{Dependence of the Fe II$_{cons}$/UV Fe II$_{60}$ and Fe II$_{incons}$/Fe II$_{cons}$ ratios of the $R_{Edd}$ (for the subset of 220 objects with FWHM Fe II$_{cons}$<5000 km/s). Dots show the mean values of the Fe II$_{cons}$/UV Fe II$_{60}$ and Fe II$_{incons}$/Fe II$_{cons}$ for objects within different ranges of log$R_{Edd}$: (1) logR$_{Edd}$<-1  (14 objects), (2) -1<logR$_{Edd}$<-0.75 (40 objects), (3) -0.75<logR$_{Edd}$<-0.5 (79 objects), (4) -0.5<logR$_{Edd}$<-0.25 (48 objects), and (5) logR$_{Edd}$>-0.25 (39 objects). The error bars are the dispersions in each
subsample.}
   \label{fig06}
\end{figure}

\subsection{Error estimates}\label{Sec3.4}

To estimate the parameter uncertainties, we performed the Monte Carlo method. First, we measured the S/N in the continuum in the 5090-5110 \AA \ range for all spectra from the sample, as the ratio of the mean value of the flux continuum to the standard deviation of the flux in the same range. We found that the average value of the S/N for the total sample is $\approx$ 35.
 Then, we divided the sample into two subsamples, with FWHM of total broad H$\beta$ (ILR+VBLR components) larger and smaller than 4000 km/s, since spectra in these two subsamples have significantly different properties \citep[see][]{Marziani2001,Marziani2018,Popovic2023}, which can affect the accuracy of the fitting procedure. We normalised the spectra using the estimated continuum level at 5100 \AA, and then we calculated the mean spectra for these two subsamples.  The model spectra for these subsamples are generated as the best fit of the mean spectra. Afterwards, we generated 100 mock spectra for each subsample, by adding the random noise to the model spectra. The random noise of mock spectra is limited in such a way that the ratio of the mean continuum level in 5090-5110 \AA \ range and $\sigma$ of noise is $\approx$ 35. After we obtained the
fitting parameters of the mock spectra, we took the 1$\sigma$ dispersion of the parameters as the parameter uncertainty.
For spectra with FWHM H$\beta$ < 4000 km/s, we estimated that uncertainties for all measured line widths are $\sim$ 3\% (FWHM H$\beta$, FWHM Fe II$_{cons}$, and FWHMs of OL, H, G+ and P+ inconsistent Fe II line groups). The uncertainties of the fluxes of Fe II$_{cons}$ and Fe II$_{incons}$ are estimated to be $\sim$ 3\% and $\sim$ 5\%, respectively. For spectra with FWHM H$\beta$ > 4000 km/s, the uncertainties of FWHM H$\beta$ and FWHM Fe II$_{cons}$ are estimated to be $\sim$ 3\%, FWHM P+ $\sim$ 5\%, FWHM G+ $\sim$ 8\%, while H and OL inconsistent Fe II groups are not present in the mean spectrum. The uncertainties of the fluxes of Fe II$_{cons}$ and Fe II$_{incons}$ are estimated to be $\sim$ 5\%. The error of continuum luminosity at 5100\AA \ is estimated to be $\sim$ 2\% for the total sample. 

The uncertainties of measurements in the UV range are estimated similarly as for the optical range. We normalised all spectra in the available UV range (338 spectra) to the estimated continuum level at $\sim$3040 \AA, and we calculated the mean spectrum in the 2650-3060 \AA \ range.  Afterwards, we generated 100 mock spectra by adding the random noise to the model spectrum, which is obtained from the best fit of the mean spectrum. The random noise is taken to be the average value of noise in the 3040–3060 \AA \ range for the total UV sample. We estimated the uncertainty of flux of  UV Fe II multiplet 60 to be $\sim$ 5\%  and the uncertainty of the FWHM of UV Fe II lines to be  $\sim$ 8\%.

\section{Results} \label{Sec4}

\subsection{Appearance of inconsistent Fe II lines in spectra}\label{Sec4.1}

 The sample of 1046 spectra was fitted with the complex Fe II template, which allows analysis of the behaviour of the different Fe II line groups. First, we analysed the presence of the diverse Fe II groups in the spectra. In 24 objects from the sample, we found no iron lines in the 4000-5600 \AA \ range, and these objects are not included in further analysis. In the rest of the 1022 objects, the iron lines are present in the optical range, but not all Fe II line groups. Namely, consistent Fe II line groups (F, S, and G) are present in almost all of the 1022 spectra (F is present in 99.5\%, S in 97\%, and G in 91\% of objects), while it is not the case for inconsistent Fe II line groups (P+, G+, OL, and H). All inconsistent line groups are present in only 272 spectra (26.6\% of the sample), while in the rest of the sample, some of these groups, or all of them, are missing. The examples of spectra with different presences of inconsistent Fe II line groups are shown in Fig. \ref{fig02_2}. The least present among inconsistent lines is the H line group, which is observed in only 32\% of spectra (marked with orange colour in Fig. \ref{fig02_2}, panel A). The OL group is observed in $\sim$50\% of the sample, while the G+ group is observed in 54\% of the sample (see blue and green lines in Fig. \ref{fig02_2}, panels A and B). The P+ line group is the most present among inconsistent line groups, and it is observed in 79\% of spectra (see red lines in Fig. \ref{fig02_2}, panels A, B, and C). In 183 objects (18\% of the sample), the inconsistent lines are completely missing, i.e. all visible Fe II lines belong to consistent Fe II line groups (see Fig. \ref{fig02_2}, panel D).
 
 We investigated  whether the appearance of the inconsistent Fe II line groups in spectra depends on some spectral properties. Therefore, we made a set of subsamples with different ranges of widths of consistent Fe II lines (see Fig. \ref{fig03}, left panels). We did the same, just for different ranges of the broad H$\beta$ line width (see Fig. \ref{fig03}, right panels). We found that all inconsistent line groups are the most present in spectra with narrower broad emission lines (FWHM<2500 km/s), where they can be observed in 70-90\% of objects. As spectra have broader emission lines, all inconsistent line groups are less present (see Fig. \ref{fig03}), but some of them disappear more effectively in spectra with broader lines than others. For example, in the subsample with 2500 km/s<FWHM Fe II$_{cons}$<5000 km/s, the inconsistent H group of lines is present in only 20\% of objects, while OL and G+ groups are present in 50-60\% of objects, and P+ in 85\% of objects. For the subsample with FWHM Fe II$_{cons}$>5000 km/s, G+, OL, and H inconsistent Fe II groups almost disappear, which is also well demonstrated in Fig.  \ref{fig04}. Exception is the P+ group, which is present even in the broadest spectra (FWHM Fe II$_{cons}$ > 8000 km/s) in 40\% of the subsample. The same trend could also be seen in examples of spectra given in Fig. \ref{fig02_2}, where the spectrum shown in panel A has the narrowest and in panel D the broadest emission lines.
  
 \begin{table}
\begin{center}
\caption{Principal component analysis of the subsample with present UV and optical Fe II lines and FWHM Fe II$_{cons}$ < 5000 km/s (220 objects). \label{T01}}

\begin{tabular}{|l|c c c|}

\hline\hline
\footnotesize{}&Comp.1&Comp.2&Comp.3    \\
\hline
 Standard deviation   & 1.75&1.19&1.10 \\
Proportion of Variance &0.44&0.20& 0.17\\
Cumulative Proportion  &0.44& 0.64& 0.81\\
 \hline
log$\lambda L_{5100}$ &0.14 & 0.02 & 0.86\\
log$R_{Edd}$ & 0.42 & -0.46 & 0.33\\
FWHM H$\beta$      &-0.36&  0.54& 0.21\\
FWHM Fe II$_{cons}$& -0.38& 0.11&  0.16\\
Fe II$_{cons}$/UV Fe II$_{60}$ & 0.35 &  0.56 &0.10\\
Fe II$_{incons}$/Fe II$_{cons}$ &0.42 &0.08&  -0.24\\
Fe II$_{incons}$/UV Fe II$_{60}$ & 0.48&  0.41 &-0.10\\

  \hline\hline

\end{tabular}
\end{center}
\end{table}

 Furthermore, we investigated whether  is disappearance of the inconsistent Fe II lines with the increase of FWHM of the Fe II linked to the effect of smearing out or to intrinsic atomic processes. Namely, we tested do inconsistent Fe II lines indeed changing their relative intensity comparing the consistent Fe II lines, or they just become indistinguishable from the continuum with increase of FWHM Fe II. For that purpose, first we compared the spectra of two well investigated objects, I Zw 1 and Mrk 493, which are both with narrow emission lines, and continuum level corresponds to the standard continuum windows. In Fig. \ref{fig03_1} we compared the Fe II template constructed on the basis of relative intensities of Fe II lines from Mrk 493, given in \cite{Park2022}, and Fe II template made from I Zw 1 spectrum by removing all other lines than Fe II \citep{Boroson1992}. We normalised both templates to the consistent line $\lambda$4520 \AA \ in order to compare do relative intensities of the other Fe II lines match in these two objects. We found a good match for all the consistent lines that do not overlap with the inconsistent lines (lines between 4500\AA \ - 4600 \AA \ from F group and 4923.9 \AA, 5018.4 \AA, 5169.0 \AA \ from S group), while in the largest part of the spectral range where there are inconsistent lines (denoted as dashed lines in Fig. \ref{fig03_1}, panels B and C), there is mismatch (see shaded regions in Fig. \ref{fig03_1}). In most of the cases, Mrk 493 Fe II template has smaller inconsistent lines than I Zw 1 object, which is the most pronounced in the red part of Fe II spectrum (G+ and H groups). The intensity of the inconsistent H group (at $\sim$5535 \AA) in I Zw 1 is almost twice as the same in Mrk 493.
 
Although these two objects have approximately similar widths of the Fe II lines, the difference between them is in almost five times larger Eddington ratio in I Zw 1, and also in significantly stronger intrinsic reddening in I Zw 1 comparing to Mrk 493 \citep[see][]{Park2022}. Therefore, we tested could intrinsic reddening be the reason of mismatch of these two spectra in red part of the Fe II bump. We used the estimated value for intrinsic reddening of I Zw 1 of E(B-V)$\sim$ 0.2 \citep{Laor1997}, to remove this effect. Then, we scaled the intrinsic reddening corrected spectrum of I Zw 1 in order to match in $\lambda$4520 \AA \ line with the non-corrected template of I Zw 1 \citep{Boroson1992} and Mrk 493 template \citep{Park2022}. The I Zw 1 template corrected for intrinsic reddening is shown in Fig. \ref{fig03_1}, panel C, denoted with the blue line. As it can be see, the effect of intrinsic reddening to intensities of the Fe II lines in red bump is small, and cannot explain significantly smaller intensities of Mrk 493 inconsistent lines comparing to the same in I Zw 1.

With this comparison, we demonstrated that the intensities of the inconsistent lines indeed vary relative to the consistent ones in different objects. However, the FWHM of Fe II lines in both compared objects is narrow, and our model gives FWHM Fe II $\approx$ 1200 km/s for both objects. To check whether inconsistent lines become smaller relative to the consistent lines for larger widths of Fe II, we performed a test. We fitted spectra with different widths of Fe II lines with both Mrk 493 and I Zw 1 templates while trying to see 
 which template fits better inconsistent lines in spectra with narrow or broad Fe II lines. Examples of the fits are shown in 
 Appendix \ref{C}, where we demonstrate that the Fe II template with stronger inconsistent lines (I Zw 1 template, \cite{Boroson1992}) provides a better fit to the spectra with narrow Fe II lines, while the Fe II template with smaller inconsistent lines (Mrk 493 template, \cite{Park2022}) gives a better fit to the spectra with broader Fe II lines. This implies that inconsistent lines decrease relative to the consistent lines with increasing the line width rather than smearing out as a pseudocontinuum.

\subsection{Searching for imprints of atomic processes}\label{Sec4.2}

To get a better understanding of the atomic processes responsible for the emission of inconsistent Fe II lines, we plot the sample in log$\lambda L_{5100}$ vs. logFWHM Fe II$_{cons}$ parameter space (see Fig. \ref{fig04}), where the objects with no H, OL, and G+ inconsistent line groups, and those in which at least one of these inconsistent line groups appears, are assigned different colours in the plot. We found that the log$\lambda L_{5100}$ vs. logFWHM Fe II$_{cons}$ relationship is different for these two groups of objects. In the case of objects with no H, OL, and G+ inconsistent line groups, a significant correlation is present (blue dots, r=0.42, P=0), while in the other group, no correlation is observed (red dots, r=0.14, P=6E-4).  We note that delimitation between these two groups of objects is approximately FWHM Fe II$_{cons}$ = 5000 km/s.

We included in the analysis the UV Fe II lines as well and compared the ratios between UV Fe II lines and consistent and inconsistent optical Fe II lines. In this way, we wanted to check whether the same processes that are responsible for the increase of the optical Fe II relative to the UV Fe II could be involved in the emission of the inconsistent optical Fe II lines and their increase/decrease relative to consistent Fe II.  Since appearance of inconsistent Fe II lines strongly depends on FWHM Fe II$_{cons}$ as shown in Figs. \ref{fig03} and \ref{fig04}, we check the dependence between the width of Fe II$_{cons}$ and ratios Fe II$_{incons}$/Fe II$_{cons}$ and Fe II$_{cons}$/UV Fe II$_{60}$. The relations are shown in Fig. \ref{fig05}. It could be seen that Fe II$_{incons}$/Fe II$_{cons}$ ratio is in significant anti-correlation with FWHM Fe II$_{cons}$ (Spearman coefficient of correlation r = -0.60, P=0), for total sample where inconsistent Fe II lines are present (838 objects). However, in the case of the Fe II$_{cons}$/UV Fe II$_{60}$ ratio, the relationship with FWHM Fe II$_{cons}$ seems to be more complex. This ratio decreases with the grow of FWHM Fe II$_{cons}$ until approximately FWHM Fe II$_{cons}$ = 5000 km/s, which is the turning point. For FWHM Fe II$_{cons}$ > 5000 km/s, it starts to increase. This relationship is shown for all objects in sample with present UV Fe II lines (338 objects). 

Since for FWHM Fe II$_{cons}$ < 5000 km/s these two ratios both showing anti-correlation with FWHM Fe II$_{cons}$, and on the other hand G+, OL, and H inconsistent Fe II line groups disappear for FWHM Fe II$_{cons}$ > 5000 km/s (see Fig. \ref{fig03}), we focused on the subset with FWHM Fe II$_{cons}$ < 5000 km/s.

We performed the PCA (see Table \ref{T01}) for the subsample with present UV and optical Fe II lines,  but with FWHM Fe II$_{cons}$ < 5000 km/s (220 objects). The parameters used in analysis are log$\lambda L_{5100}$, log$R_{Edd}$, width of consistent optical Fe II lines, width of the broad H$\beta$ line, and ratios Fe II$_{cons}$/UV Fe II$_{60}$, Fe II$_{incons}$/Fe II$_{cons}$ and Fe II$_{incons}$/UV Fe II$_{60}$. The first eigenvector describes 44\% of the variance. It is dominated by the width of the consistent Fe II lines and the Eddington ratio. It could be seen that as the widths of the broad H$\beta$ and Fe II$_{cons}$ decrease, $R_{Edd}$ increases, and the intensity of the optical Fe II increases relative to the UV Fe II, as well as the intensity of the optical Fe II$_{incons}$ relative to the Fe II$_{cons}$. The influence of the continuum luminosity in eigenvector 1 is negligible.  In the second eigenvector, which describes 20\% of variance, the dominant physical parameter is the FWHM H$\beta$, which increases together with Fe II$_{cons}$/UV Fe II$_{60}$. In the third eigenvector, the dominant physical parameter is the continuum luminosity.

The PCA implies the importance of the two physical parameters in understanding the processes of the Fe II emission.  These are the width of Fe II$_{cons}$ and $R_{Edd}$, which dominate in eigenvector 1. Therefore, we performed correlations between these parameters and different Fe II ratios (see Table \ref{T02}) for the same subset used for PCA (220 objects with present UV Fe II lines and FWHM Fe II$_{cons}$ < 5000 km/s). As it can be seen in Table \ref{T02}, the Fe II$_{cons}$/UV Fe II$_{60}$, Fe II$_{incons}$/Fe II$_{cons}$  and Fe II$_{incons}$/UV Fe II$_{60}$ ratios show a growing trend for narrower Fe II lines, and higher Eddington ratio. The correlations between optical-to-UV FeII ratio vs. FWHM Fe II$_{cons}$ and $R_{Edd}$ become more significant when observed only for Fe II$_{cons}$ G group.
 We illustrated the growing trend of Fe II$_{incons}$/Fe II$_{cons}$ and Fe II$_{cons}$/UV Fe II$_{60}$ ratios with $R_{Edd}$ in Fig. \ref{fig06}, where the values of both ratios are binned for objects within different log$R_{Edd}$ ranges.  The correlation between Fe II$_{incons}$/Fe II$_{cons}$ and Fe II$_{cons}$/UV Fe II$_{60}$ ratios in this subset is r = 0.30, P=6E-6.

\begin{table}
\begin{center}
\caption{Correlations between the different Fe II line ratios and some physical parameters.  
\label{T02}}

\begin{tabular}{| c | c |c |c |}
\hline

&  & FWHM Fe II$_{cons}$ &  logR$_{Edd}$  \\
\hline
 \hline
    \multirow{2}{*}{ \large $\frac{\rm Fe II_{\it cons}}{\rm UV Fe II_{60} }$ } &r  & -0.22  &0.21\\
    &P  &  8E-4  &0.001\\

\hline

   \multirow{2}{*}{ \large $\frac{\rm Fe II_{\it cons} G}{\rm UV Fe II_{60} }$ }&r  &  -0.31& 0.25\\
    &P  &  2E-6& 2E-4 \\
\hline
 \multirow{2}{*}{\large $\frac{\rm Fe II_{\it incons}}{\rm UV Fe II_{60} }$ }&r   & -0.40 &   0.35  \\
    &P  &  4E-10 &  6E-8\\
 \hline

\multirow{2}{*}{\large $\frac{\rm Fe II_{\it incons}}{\rm Fe II_{\it cons} }$ }&r   & -0.43&  0.37 \\
    &P  &  3E-11  &  1E-8\\
 \hline

\hline

\end{tabular}
\tablefoot{The correlations are done for the subsample of 220 objects used in the PCA (with present UV and optical Fe II lines and FWHM Fe II < 5000 km/s). The table contains Spearman coefficients of correlation (r) and P-values. 
}

\end{center}
\end{table}

\subsection{Differences between the widths of the Fe II line groups}\label{Sec4.3}

\begin{figure*} 
\centering
 \includegraphics[width=183mm]{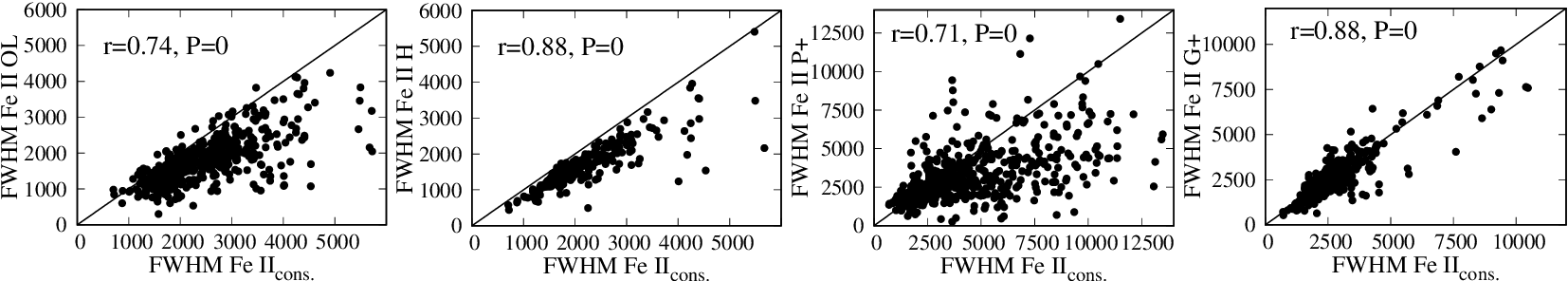}
 \caption{Comparison of the widths of the different inconsistent Fe II line groups with the widths of the Fe II$_{cons}$. The Spearman coefficients of correlation (r) and P-values are given in graphs. The solid line represents a 1-to-1 relationship. The widths are given in km/s.}
   \label{fig07}
\end{figure*}

 We compared the widths of the Fe II$_{cons}$ with the widths of the inconsistent Fe II line groups, and UV Fe II. 
Using the subsample of 338 objects, where both optical and UV Fe II lines are present, we found the weak correlation between the FWHMs of Fe II$_{cons}$ and UV Fe II lines (r = 0.37, P = 9E-12), which is in accordance with results obtained in \cite{Kovacevic2015}. Also, we found that the mean values of widths of UV and optical Fe II lines are very similar (FWHM Fe II$_{cons}$ = 4740 km/s, FWHM UV Fe II = 4810 km/s).  On the other hand, we found that the widths of all inconsistent line groups  are in strong correlation with FWHM Fe II$_{cons}$, although their values do not follow a 1-to-1 relationship in all cases (see Fig. \ref{fig07}). The widths of the Fe II$_{incons}$ OL and Fe II$_{incons}$ H are smaller than the widths of the Fe II$_{cons}$ for all objects. The widths of the Fe II$_{incons}$ P+ follow a 1-to-1 relationship up to $\approx$4000 km/s, and then they become  smaller than the widths of the Fe II$_{cons}$. Only the widths of the Fe II$_{incons}$ G+ lines approximately follow a 1-to-1 relationship with FWHM Fe II$_{cons}$ even for the larger widths. The average values of the widths of different Fe II line groups are given in Table \ref{T04}. The a\-ve\-ra\-ge widths are calculated for different subsamples, formed following the presence of particular inconsistent line groups. The smallest subsample consists of the spectra where all inconsistent lines are present (H, OL, G+, and P+), while the other two subsamples consist of the spectra where OL, G+, and P+ are present and the spectra with present P+ lines. Since P+ inconsistent lines are the most present in spectra among inconsistent line groups (see Fig. \ref{fig03}), this subsample is the largest. It could be seen that H and OL have similar average values of the widths, and they are smaller than the widths of the G+, P+, and Fe II$_{cons}$. The G+ and P+ have similar average values of the widths as Fe II$_{cons}$ for subsamples where both of these two line groups are present. However, in the largest subsample, which consists of the spectra with present P+ lines (811 spectra), which includes spectra with very broad lines (see Fig. \ref{fig02_2}, panels C and D), the average width of the P+ lines is smaller than the average width of the consistent Fe II lines.

\begin{table}
\begin{center}
\caption{Average values of the FWHMs of different Fe II line groups calculated for different subsets. \label{T04}}
\begin{tabular}{|c | c | c |c |}
\hline

  & \multicolumn{3}{c|}{ {\small average value $\pm$ SD [km s$^{-1}$]} } \\
  \hline 
 & {\small subset (272)} & {\small subset (498)} & {\small subset (811)}\\
  &{\small H, OL, P+, G+ }& {\small OL, G+, P+ }& {\small P+}\\
\hline 
\hline
{\small FWHM Fe II$_{cons}$} &{\small 2040$\pm$690 } &{\small 2315$\pm$900 } & {\small 3720$\pm$2500 }\\
 \hline
{\small FWHM Fe II P+} & {\small 2000$\pm$560 }& {\small 2290$\pm$930} & {\small 2860$\pm$1620}\\
 \hline
{\small FWHM Fe II G+} &{\small 1990$\pm$710 } & {\small 2200$\pm$915} & -- \\
 \hline
 {\small FWHM Fe II OL} &{\small 1560$\pm$520 } & {\small 1710$\pm$680}& --   \\
  \hline
 {\small  FWHM Fe II H} & {\small 1590$\pm$490 } & --& --\\
   \hline
 {\small  FWHM H$\beta$ }& {\small 2890$\pm$890 }& {\small 3100$\pm$1100}&{\small 3590$\pm$1445}\\

\hline

\end{tabular}

\tablefoot{The subset with present all inconsistent Fe II line groups is given in the second column (272 objects), and subsets with present some of the inconsistent Fe II line groups are given in the third column (present OL, G+ and P+ line groups, 498 objects) and in the fourth column (present P+ line group, 811 objects).
}
\end{center}
\end{table}

\section{Discussion} \label{Sec5}

To understand complex Fe II emission in AGN spectra, with special focus on the inconsistent Fe II lines, we divided optical Fe II emission into several line groups according to the atomic properties and similar behaviour in spectra and analysed their relationship with UV Fe II and some measured spectral properties. In this section we discuss the obtained results from se\-ve\-ral aspects, trying to give a possible explanation for the observed Fe II properties.

The results obtained in Sect. \ref{Sec4.2} implies that increase of Eddington ratio is followed with spectral narrowing, an increase of the optical Fe II lines relative to UV Fe II lines and, at the same time, an increase of inconsistent optical Fe II lines relative to consistent ones (see Tables \ref{T01} and \ref{T02} and Figs. \ref{fig05} and \ref{fig06}). 
However, the entire system that emits the optical and UV Fe II lines appears to be very complicated. While mentioned correlations are reflected in the first eigenvector of PCA, which dominates over the variance of the sample (44\%), in the second eigenvector of PCA, which has a significantly smaller variance (20\%), the increase of the optical Fe II lines relative to the UV Fe II is followed by increase of the width of H$\beta$, and the decrease of the Eddington ratio. In this case consistent and inconsistent optical Fe II lines increase at the same rate relative to the UV Fe II i.e., the Fe II$_{incons}$/Fe II$_{cons}$ ratio does not change. Also, the Fe II$_{cons}$/UV Fe II$_{60}$ ratio has complex relationship with the width of Fe II lines. It decreases as the width of Fe II lines increases up to $\approx$ 5000 km/s, and than it starts to grow (see Fig. \ref{fig05}).

  If we apply a more detailed analysis of inconsistent Fe II lines,  and separate them into line groups, we may notice that there are some differences between them. As the widths of lines in spectra are larger, the Fe II$_{incons}$ H group disappears first, then G+ and OL, and finally the Fe II$_{incons}$ P+ lines. The H, G+, and OL line groups can hardly be seen for FWHM Fe II > 5000 km/s, while the P+ inconsistent line group can be visible in the spectra with broader lines.

The presented results  open several interesting questions regarding the emission of the Fe II lines. The basic question is what atomic processes causes the strong emission of inconsistent Fe II lines. We tried to understand why these lines are significantly stronger than expected by theoretical calculations using their atomic properties. It is also questionable why they disappear in spectra with broader emission lines, and why they do not disappear all at once.

\subsection{Process of redistribution of photons from stronger to weaker transitions}\label{Sec5_2}

According to atomic data, the inconsistent optical Fe II lines should be up to two order of magnitude smaller than consistent Fe II, i.e., they should not be visible in spectra at all. However, in our sample, in only 18\% of spectra with present iron lines, none of the inconsistent Fe II line groups were seen. In all other spectra, at least some of the inconsistent line groups were observed, while in $\sim$27\% of our sample, all inconsistent Fe II lines are present, and they have intensities just slightly smaller than the intensities of the consistent Fe II lines. 

Similarly, following the atomic data, it is expected that optical Fe II emission lines should be several orders of magnitude smaller than UV Fe II. However, the UV and optical Fe II emissions are of the same order of magnitude in the majority of spectra \citep{Joly1981}. We found an analogy between this case and the relative intensities of the Fe II$_{cons}$ and Fe II$_{incons}$ lines. This motivated us to check whether the same processes of energy transfer from stronger to weaker lines could be responsible for both of these phenomena.

Indeed, the idea that the same process might be triggering the growth of both of these ratios is supported by some results (see EV1 in Table \ref{T01} and Fig. \ref{fig06}). The problem of the optical Fe II/UV Fe II ratio and the excess of the optical Fe II emission was the subject of many previous studies \citep{Joly1981, Collin2000, Lu2019}. One of the widely accepted explanations of this phenomenon is the atomic process given in the literature as the radiation trapping effect \citep[see also optical depth effect, radiation imprisonment,][]{Holstein1947, Molisch1992}. This is the process of self-absorption, which becomes significant if, during the radiative transfer, the optical depth of the line is high, which brings to saturation of its intensity. This process could be described as a reduction in spontaneous emission as: $$A'=\epsilon\cdot A,$$ where $A$ is the Einstein coefficient for spontaneous emission, $A'$ is the effective spontaneous transition probability, and $\epsilon$ is the mean/local escape probability, where $\epsilon <<$ 1 
\citep{Phillips1978b, Irons1979, Netzer1980,Kwan1981, Verner1999, Collin2000}. 

\begin{figure} 
\centering
  \includegraphics[width=65mm]{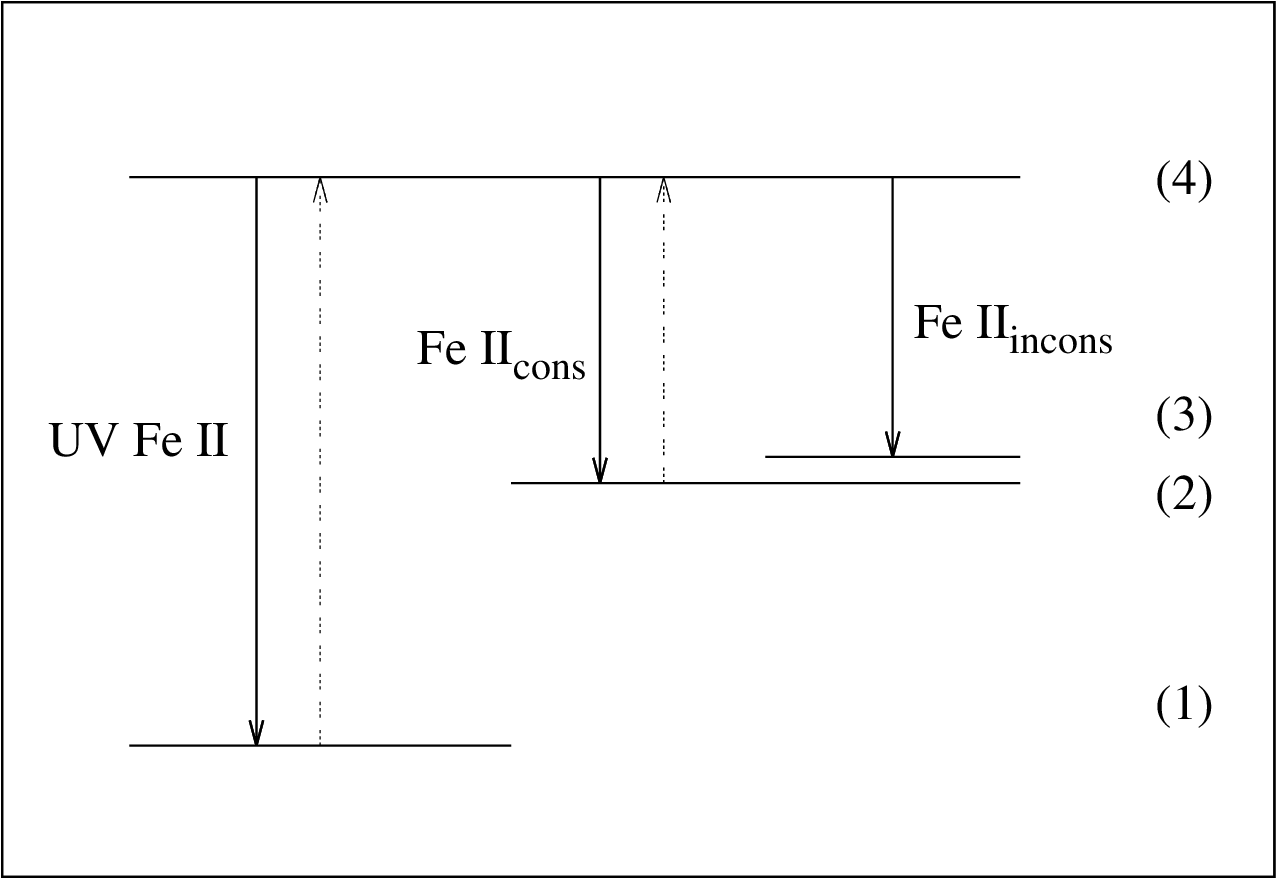}

   \caption{Simplified Grotrian diagram of Fe II with four active energy states. The UV Fe II transitions are denoted as 4-1, optical transitions of consistent Fe II lines as 4-2 and optical transitions of inconsistent Fe II lines as 4-3.}
    \label{fig08}
\end{figure}

To illustrate the possible energy transfer from UV to optical Fe II emission, we used the simplified Grotrian diagram of the Fe II ion given in Fig. \ref{fig08}, with four active levels: (1) ground metastable state, (2) and (3) middle metastable states, and (4) common upper level for UV and optical transitions in Fe II. The UV Fe II emission arises in 4 - 1 transitions, and optical emission arises in 4 - 2 (consistent Fe II lines) and 4 - 3 (inconsistent Fe II lines). 
In this subsection we focus only on UV and consistent optical Fe II lines. For UV Fe II lines, the Einstein coefficient $A$ for spontaneous emission is $\sim$10$^8$ s$^{-1}$, while for optical emission of consistent lines, it is $\sim$10$^5$ s$^{-1}$. Therefore, spontaneous emission from upper level (4)  has significantly larger probability for UV (4 - 1), than for optical transitions (4 - 2). However, a larger probability for spontaneous emission ($A$) leads to a larger Einstein coefficient for absorption ($B$) because they are connected with the following relation:

$$Aji=8\pi h/\lambda_{ij}^3\cdot g_i/g_j\cdot Bij,$$
 where $\lambda$ is the wavelength of transition, and $g_i$ and $g_j$ are statistical weights. This means that UV Fe II lines have a higher probability of emission, but also a higher probability of absorption than optical Fe II emission lines. High spectral energy density in UV and a high population of electrons in lower energy states lead to multiple absorptions of UV photons and an effective reduction of the coefficient for spontaneous emission to $A'=\epsilon\cdot A$. This causes a decrease in the UV line intensity. After multiple emissions and absorptions of the UV photons, their optical depth grows, i.e., the escape probability $\epsilon$ decreases up to 10$^{-3}$, which brings an efficient Einstein coefficient for spontaneous emission of UV photons ($A'$) of the same order of magnitude as for optical Fe II emission. In other words, if an electron is in an excited state (4), it has an equal probability of being deexcited to level (1) or to level (2), i.e., to be emitted as UV or as an optical photon. In this way, the UV transitions are resonantly pumping the optical transitions and producing the transfer of energy of radiation from UV to the optical part of the spectra \citep[see][]{Netzer1980, Joly1981, Collin2000}.

 \subsection{Cause of the increase of the inconsistent Fe II lines }\label{Sec5_3}

Taking into account the analysis of the process of self-absorption described in Sect. \ref{Sec5_2}, which leads to the redistribution of the photons from transitions of higher probability (UV Fe II) to transitions of lower probability (consistent optical Fe II), it would be interesting to consider if this mechanism could also have an important role in the violation of expected relative intensities between consistent and inconsistent optical Fe II lines. The Einstein coefficient for spontaneous emission of inconsistent lines in G+, H, and OL line groups is $\sim$10$^3$ s$^{-1}$, while for consistent Fe II lines it is $\sim$10$^5$ s$^{-1}$. Therefore, the inconsistent Fe II lines from these groups have a two order-of-magnitude lower Einstein coefficient for spontaneous emission than consistent Fe II, which means that they should not be visible in spectra. However, their intensities are of the same order of magnitude as the intensities of the consistent Fe II lines in large amounts of spectra. The only exception among inconsistent line groups is P+, whose Einstein coefficient for spontaneous emission is of the same order of magnitude as for consistent lines ($\sim$10$^5$ s$^{-1}$), and it is analysed in the next section. 

When electrons populate the lower level (2) of strong optical transitions (consistent Fe II lines), they cannot deexcitate to the level (1) because it is a forbidden transition. Therefore, their lower level (2) is metastable, and they could be self-absorbed. With an increase in the radiation density of these lines, their absorption probability increases. After multiple emissions and absorptions, the escape probability  $\epsilon$ decreases, i.e., their efficient coefficient for spontaneous emission  ($A'$) becomes in order of magnitude of the Einstein coefficient for spontaneous emission of the weak optical emission lines (inconsistent Fe II lines). On the other hand, the weak optical Fe II lines have small optical depths and have negligible self-absorption, so finally both consistent and inconsistent Fe II lines could be seen in spectra, and their intensities are of the same order of magnitude. In the proposed scenario, in the presence of certain physical conditions, the process of radiation trapping is triggered, and the consistent Fe II lines are resonantly pumping the much weaker, inconsistent Fe II lines, enabling in this way their appearance in spectra.

\subsection{Specific case of Fe II$_{incons}$ P+ line group}\label{Sec5_4}

Considering Fe II$_{incons}$ P+ lines, there are two interesting phenomena. The first is that regardless of the fact that the strongest P+ lines (4173.5 \AA, 4233.2 \AA) have Einstein coefficients for spontaneous emission of $\sim$10$^5$ s$^{-1}$, which is similar to consistent lines, they are less present in spectra with very broad lines than Fe II$_{cons}$. They could be seen in only 40\% of the spectra with FWHM Fe II$_{cons}$ > 8000 km/s, in which consistent Fe II lines are present. The second interesting thing considering these lines is  when they are present in spectra, some lines within P+ group (e.g. 4128.7 \AA, 4296.6 \AA, 4303.2 \AA),  which should have significantly smaller intensity compared to the strongest P+ lines, are approximately equally strong as these lines, which means that some processes violate their relative intensities. 

Both processes are probably connected with the self-absorption of the P+ lines. The atomic property that distinguishes P+ lines from consistent Fe II lines is that lower levels of the P+ transitions (b$^4$P for 14 lines and b$^4$F for 3 lines) have smaller energy than lower levels of the consistent Fe II transitions (F and G groups). This could make the process of self-absorption more efficient for P+ lines, which may result in a decrease in their intensity. Namely, the efficiency of photon emission in an atom is controlled by two processes: excitation to the upper energetic level and depopulation of the lower energetic level.  If processes of excitation of the electrons to the upper energetic level are efficient, and the Einstein coefficient for spontaneous emission ($A$) is high, but processes of depopulation of the lower energetic level are not efficient, this could result in 
the small intensity of the emission lines. In these cases, the population of the lower level is large, which increases the self-absorption probability, i.e., the probability of the absorption of emitted photons. It means that the number of emitted photons will be large, but at the same time, photons will be absorbed by the resonant process. It results in a smaller intensity of these lines in spectra than expected by theoretical calculations using their atomic properties. On the other hand, the process that violates relative intensities between P+ lines is probably the radiation trapping effect within P line transitions.

\subsection{Physical conditions that trigger atomic processes}\label{Sec5_5}

The radiation trapping effect becomes an efficient process in the case of some transitions if there is a large optical depth for these transitions. However, we found that Fe II$_{opt}$/Fe II UV and Fe II$_{incons}$/Fe II$_{cons}$ ratios both increase when the Eddington ratio increases and when the width of broad lines decreases. This brings one to the question of whether macroscopic properties estimated from spectra, such as R$_{Edd}$, could be connected with the physical quantity as optical depth.

\cite{Sameshima2011} also found that the Fe II$_{opt}$/Fe II UV ratio increases with an increase in the R$_{Edd}$. Assuming that the process of self-absorption is responsible for the increase of the Fe II$_{opt}$ relative to the Fe II UV \citep{Collin2000}, they suppose that the increase of the R$_{Edd}$ causes an increase in the optical depth and column density in Fe II lines. In this scenario, the small column density clouds are probably driven away from the emission region because of the large radiation pressure caused by large R$_{Edd}$, and only the large column density clouds can be gravitationally bound \citep{Dong2011, Sameshima2011}. It is possible that large radiation pressure, caused by a high level of accretion, prevents the formation of the Fe II emission (and also Balmer lines) close to the central supermassive black hole, so they arise farther away in the ILR \citep{Popovic2023}. Therefore, we observe narrower Fe II and Balmer emission lines in spectra with high R$_{Edd}$ and strong Fe II inconsistent lines.

\subsection{Comparison with Cloudy code calculations}\label{Sec5_6}

\begin{figure*} 
\centering
  \includegraphics[width=160mm]{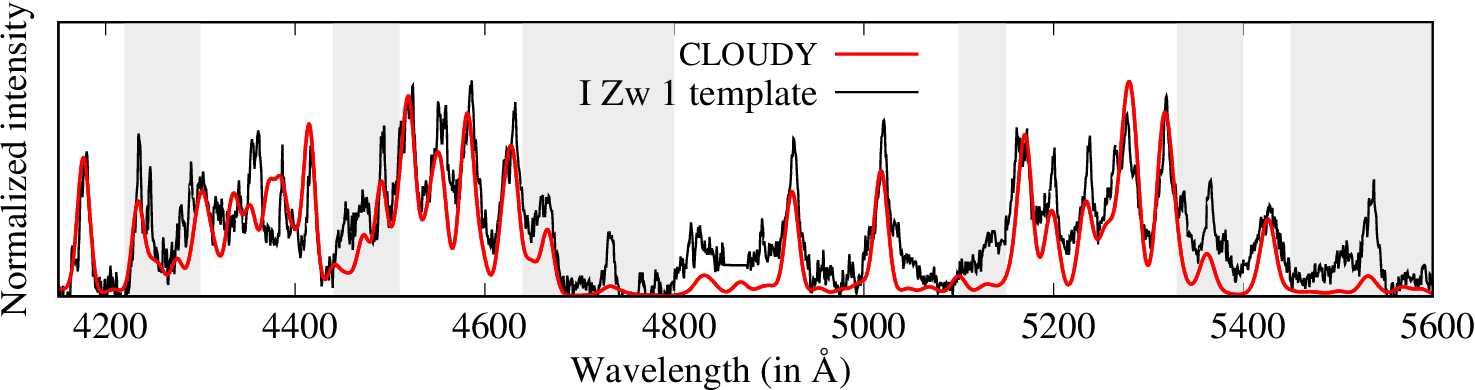}

   \caption{Comparison of the I Zw 1 template \citep{Boroson1992} with the CLOUDY theoretical tamplate (C23.00) calculated for log$\Phi_H$ = 21 (cm$^{-2}$ s$^{-1}$), log$n_H$ = 11 (cm$^{-3}$), and $v_{turb}$ = 20 kms$^{-1}$. Shaded regions highlight the inconsistent Fe II lines in  I Zw 1, which are underestimated by theoretical tamplate.}
    \label{fig09}
\end{figure*}

There were many attempts to calculate and predict the Fe II emitted spectrum in AGNs using sophisticated spectral synthesis code Cloudy, which is designed to simulate physical conditions within an astronomical plasma \citep[see][]{Bruhweiler2008, Smyth2019, Sarkar2021,Zhang2024, Pandey2024}. However, although the amount of the input atomic parameters of Fe II ion increased through the new versions of this code, the reproduction of the observed UV to optical Fe II ratio, as well as the relative intensities among the Fe II lines, still remains a challenge \citep[see][]{Zhang2024, Pandey2024}.

\cite{Zhang2024} used Cloudy C22.01 code \citep{Ferland2017} in order to obtain the physical conditions of the Fe II emission region in I Zw 1 and Mrk 493. They assumed the solar abundance and column density of 10$^{24}$ cm$^{-2}$, and used various atomic data which includes 716 levels up to 26.4 eV, and about 256000 Fe II emission lines \citep{Smyth2019, Sarkar2021}. They vary the density ($n_H$) and ionising photon ﬂux ($\Phi_H$) in order to achieve the best fit of the optical Fe II emission lines in I Zw 1 and Mrk 493, but also to reach the observed ratio of optical EW Fe II relative to the UV continuum at 1215 \AA. They found that model with log$\Phi_H$ = 20.5 (cm$^{-2}$ s$^{-1}$), log$n_H$ = 11 (cm$^{-3}$), and microturbulent velocity $v_{turb}$ = 100 kms$^{-1}$, gives the best solution for these spectra. However, they found that some Fe II lines in simulated spectrum are significantly smaller comparing to the observed  Fe II lines in I Zw 1 and Mrk 493. \cite{Zhang2024} try to check could the presence of the other atomic or molecular lines explain the missing flux, but including all available levels and emission lines from the other elements in Cloudy calculation did not change the simulated spectrum. As it could be seen in \cite{Zhang2024} (see their Fig. 2 and Fig. A3), the lines which are underestimated or completely missing in simulated spectrum comparing the I Zw 1 and Mrk 493, are the lines assigned as inconsistent in this work. The underestimated and missing lines are in ranges 4440-4510 \AA, 4640-4800 \AA, 5100-5150 \AA \ and 5330-5400 \AA, with addition of two consistent lines at 5197 \AA \ and 5234 \AA. Their flux remains underestimated even for the changing of the different input parameters (see the appendix of \citealt{Zhang2024}).

To investigate how inconsistent Fe II lines vary relative to consistent by changing the physical parameters, we used the set of the theoretical templates given in \cite{Pandey2024}, which are calculated using Cloudy C23.00 code \citep{Chatzikos2023}, for solar abundance and column density of 10$^{24}$ cm$^{-2}$. We noticed that intensity of inconsistent lines relative to consistent grows for smaller microturbulent velocities, and we focused to find the theoretical model which best describes the relative intensities among the optical Fe II lines in I Zw 1, without taking into account the ratio of UV to optical Fe II. We found that synthetic model which best describes the relative intensities of optical Fe II lines in I Zw 1 is for log$\Phi_H$ = 21 (cm$^{-2}$ s$^{-1}$), log$n_H$ = 11 (cm$^{-3}$), and $v_{turb}$ = 20 kms$^{-1}$ parameters, which is plotted in Fig. \ref{fig09}. In this model, the inconsistent lines are larger relative to consistent lines comparing to the models with higher microturbulent velocity, but still these lines are underestimated, as it can be seen in shaded regions in Fig. \ref{fig09}.

It seems that single model calculated including set of physical parameters cannot reproduce well the Fe II emission as seen in I Zw 1 from all aspects. Large microturbulent velocity is needed for achieving better fit of optical to UV Fe II ratio \citep{Zhang2024}, but it strongly underestimates the flux of inconsistent optical Fe II lines relative to consistent. Their flux slightly increases with decrease of microturbulent velocity. These results suggest that Fe II emission region is probably stratified, which is also supported by observed smaller average widths for some inconsistent Fe II lines (OL and H groups) comparing the consistent Fe II lines (see Table \ref{T04}). Following the model of locally optimally emitting clouds (LOCs) given in \cite{Baldwin1995}, it is possible that there are several Fe II emission regions in AGN structure, with different physical conditions, resulting in different relative intensities of emitted Fe II lines. In this scenario, the observed Fe II spectrum is complex mixture of radiation from LOCs positioned at different distance from the black hole. On the other hand, it is also possible that viewing angle in combination with flattened geometry of the Fe II emission region affects the observed properties of Fe II lines, especially their relative intensities \citep[see][]{Panda2019, Gaskell2022}.

\section{Conclusions}\label{Sec6}

In this research, we used the sample of 1046 Type 1 AGN spectra in order to investigate the Fe II emission lines. For this purpose, we constructed and applied the flexible and complex Fe II model, where Fe II lines are divided into several line groups and fitted independently. This approach allowed us to perform a sophisticated analysis and investigate the role of the different atomic processes in Fe II emission. The Fe II lines were divided into two large groups: consistent, those whose relative intensities are in accordance with their atomic properties (F, S, and G line groups), and inconsistent, those whose relative intensities are significantly stronger than theoretically expected following their transition probabilities (P+, G+, OL, and H line groups). We especially focused on understanding the processes that produce strong inconsistent Fe II lines, and therefore we investigated the correlations between inconsistent and consistent Fe II lines with UV Fe II lines and some measured spectral properties. Summarising the main results, we draw the following conclusions:

\begin{enumerate} 
\item The presence and strength of the inconsistent Fe II lines relative to the consistent Fe II lines are correlated with the line widths. They are all present in the case of the spectra with narrow Fe II lines (NLSy1s). As spectra have broader emission lines, the inconsistent Fe II lines have smaller intensities relative to consistent Fe II, and in the case of very broad spectra, they disappear. With the increase in line widths, the Fe II$_{incons}$ H group disappears first, then G+ and OL, and finally the Fe II$_{incons}$ P+ line group. The Fe II$_{incons}$ H, G+, and OL lines rarely exist in spectra with FWHM Fe II > 5000 km/s, while Fe II$_{incons}$ P+ could be seen in 40\% of the spectra with very broad lines (FWHM Fe II > 8000 km/s).
 
\item The ratios Fe II$_{incons}$/Fe II$_{cons}$ and Fe II$_{opt}$/Fe II$_{UV}$ both increase with an increase in the Eddington ratio. It is possible that in both cases, the process of radiation trapping (self-absorption) is responsible for the transmission of the energy from the UV to the optical Fe II emission lines and, analogous to that, from the Fe II$_{cons}$ to the Fe II$_{incons}$. In this scenario, the high Eddington ratio causes an increase in the optical depth in Fe II lines, which results in the triggering of the process of self-absorption. That process results in an increase of the optical Fe II relative to UV Fe II lines and, at the same time within the optical Fe II lines, an increase of the Fe II$_{incons}$ relative to the Fe II$_{cons}$ lines.
 
\item The reason for the absence of P+ lines in objects with large widths of Fe II lines is probably inefficient depopulation of the lower energetic level of these transitions, which leads to the large self-absorption of these lines. The P+ lines are strong in spectra with narrower lines, where emission regions probably have a larger electron density, and therefore collision depopulation of lower transition levels is more effective.

\item  There are probably several Fe II emission regions with different physical conditions and distances from the black hole that emit Fe II spectra with different relative intensities among Fe II lines. The observed Fe II spectrum is a complex mixture of radiation from these emission regions. The different average widths of some Fe II line groups also indicate the stratification of the Fe II emission region. The G+ and P+ groups have similar average values of the widths as Fe II$_{cons}$ for subsamples, where both of these line groups are present. However, the H and OL groups have smaller average values of the widths than the widths of the Fe II$_{cons}$, which probably implies that these lines arise farther away from the central engine compared to the other Fe II lines. 
    
\end{enumerate}
 
Future research should be devoted to more careful consideration of the possibility that Fe II emission region have complex structure, with different physical properties. Possible stratification of the Fe II emission region might be essential for explaining some of the poorly understood characteristics of Fe II emission lines.

\begin{acknowledgements}
We thank the anonymous referee for valuable comments and
suggestions that helped us to significantly improve the paper.
This research was supported by the Ministry of Science, Technological Development and Innovation of the Republic of Serbia through contract no. 451-03-66/2024-03/200002 and 451-03-65/2024-03/200162. 

Funding for the Sloan Digital Sky Survey IV has been provided by the 
Alfred P. Sloan Foundation, the U.S. 
Department of Energy Office of 
Science, and the Participating 
Institutions. 

SDSS-IV acknowledges support and 
resources from the Center for High 
Performance Computing  at the 
University of Utah. The SDSS 
website is www.sdss4.org.

SDSS-IV is managed by the 
Astrophysical Research Consortium 
for the Participating Institutions 
of the SDSS Collaboration including 
the Brazilian Participation Group, 
the Carnegie Institution for Science, 
Carnegie Mellon University, Center for 
Astrophysics | Harvard \& 
Smithsonian, the Chilean Participation 
Group, the French Participation Group, 
Instituto de Astrof\'isica de 
Canarias, The Johns Hopkins 
University, Kavli Institute for the 
Physics and Mathematics of the 
Universe (IPMU) / University of 
Tokyo, the Korean Participation Group, 
Lawrence Berkeley National Laboratory, 
Leibniz Institut f\"ur Astrophysik 
Potsdam (AIP),  Max-Planck-Institut 
f\"ur Astronomie (MPIA Heidelberg), 
Max-Planck-Institut f\"ur 
Astrophysik (MPA Garching), 
Max-Planck-Institut f\"ur 
Extraterrestrische Physik (MPE), 
National Astronomical Observatories of 
China, New Mexico State University, 
New York University, University of 
Notre Dame, Observat\'ario 
Nacional / MCTI, The Ohio State 
University, Pennsylvania State 
University, Shanghai 
Astronomical Observatory, United 
Kingdom Participation Group, 
Universidad Nacional Aut\'onoma 
de M\'exico, University of Arizona, 
University of Colorado Boulder, 
University of Oxford, University of 
Portsmouth, University of Utah, 
University of Virginia, University 
of Washington, University of 
Wisconsin, Vanderbilt University, 
and Yale University.
\end{acknowledgements}


\appendix

\section{Determination of the continuum level}\label{A}

In the case of objects with very broad and strong Fe II emission lines, the wings of the numerous Fe II lines, broad Balmer lines, and He II are blended, forming the pseudocontinuum, which is difficult to separate from the real continuum emission of an AGN \citep{Popovic2023}. To reduce uncertainty in the determination of continuum level, we reduced the number of free parameters for fitting the Balmer lines, by adding the fitting constraints for these lines. The broad components of these lines are fitted with the sum of the ILR and VBLR Gaussians, whose widths, shifts, and relative intensities to each other are the same for H$\beta$, H$\gamma$ and H$\delta$ lines, so these lines have the same shapes in one spectrum. Additionally, the total intensities of the broad components of these lines are fixed to follow the theoretical value for case B \citep[see][]{Osterbrock2006}. In this way, the broad components of H$\beta$, H$\gamma$ and H$\delta$ lines are described with six free parameters: width and shift of ILR component, width and shift of VBLR component, intensity of VBLR relative to ILR, and intensity of the broad H$\beta$, while intensities of broad H$\gamma$ and H$\delta$ are respectively 0.504 and 0.305 of broad H$\beta$ intensity. Additionally, in order to more correctly reproduce the Fe II pseudocontinuum, we included two-component model for the Fe II lines that belong to the consistent lines.

To correctly determine the continuum level, it is important to include all lines in the analysed range, which may influence the fitting procedure. Therefore, we added in the fitting model the broad He I 4027.3 \AA \ line, which is fitted with single Gaussian, and doublet [S II] 4069.7, 4077.5 \AA \ lines. The [S II] lines are fitted with two narrow Gaussians and have the same width and shift as all other narrow lines in the analysed range ([O III] 4959, 5007 \AA \ lines, [O III] 4363 \AA \ and narrow Balmer lines).   Finally, the continuum is modelled with a power law that is simultaneously fitted with the emission lines, and the only adopted continuum window is near $\sim$ 5600 \AA \ \citep{Tsuzuki2006}. 

\section{Properties of the complex Fe II template}\label{B}

 In \cite{Kovacevic2010}, it is assumed that all Fe II lines have only one component, and they were modelled with single Gaussians in order to make a simplified Fe II template, which would be convenient for use. \cite{Popovic2023} modelled the large set of AGN synthetic spectra and found that broad H$\beta$ and Fe II lines could be well modelled as the sum of the emission from two BLR sub-regions, ILR and VBLR. Therefore, in this modified version of the Fe II template,  we include two Gaussian components for the Fe II lines that belong to the consistent lines. We assign these components as Fe II ILR and Fe II VBLR components, assuming that the Fe II emission arises from different layers of the BLR. 

By careful analysis of several high S/N spectra with narrow and well defined Fe II lines we found that consistent Fe II lines are generally broader than inconsistent, and that best fit is achieved with model of double-Gaussian for consistent Fe II lines and single-Gaussian component for inconsistent lines. This approach is supported by empirical clues from several studies of optical Fe II lines \citep{Veron-Cetty2004, Dong2008, Bruhweiler2008, Dong2011, Marinello2016,Park2022}, who noticed that Fe II lines have two components: a broader and narrower one, but also that some Fe II lines have only narrow component. 

 \cite{Veron-Cetty2004} analysed the Fe II lines in I Zw 1, while \cite{Park2022} analysed the Fe II lines in Mrk 493, and both concluded that optical Fe II lines could be divided into two groups: broader and narrower ones. For example, in both studies, they found that some lines in multiplets 27, 28, 37 and 49 (Fe II $\lambda$4173 \AA, Fe II $\lambda$4233 \AA, Fe II $\lambda$4629 \AA \ and Fe II $\lambda$5425 \AA), which belong to P+ and G+ inconsistent line groups in our model, have only narrow component of Fe II, or narrow component is dominant. \cite{Bruhweiler2008} and \cite{Marinello2016} also found that optical Fe II lines should be modelled with narrow and broad component, but they emphasise that some Fe II lines have only the narrow component. \cite{Bruhweiler2008} found that Fe II lines in 4200-4500 \AA \ range are narrower than Fe II lines in 4500-4700 \AA \ range, and arise in low density region. These lines in our model belong to P+ and OL inconsistent line groups. Generally, the inconsistent Fe II lines from this work are  defined as narrower or with dominant narrow component in several previous empirical studies. Therefore, they are fitted with single Gaussian model, with free parameter for the width for each inconsistent line group.

\begin{figure} 
\centering
 \includegraphics[width=69mm]{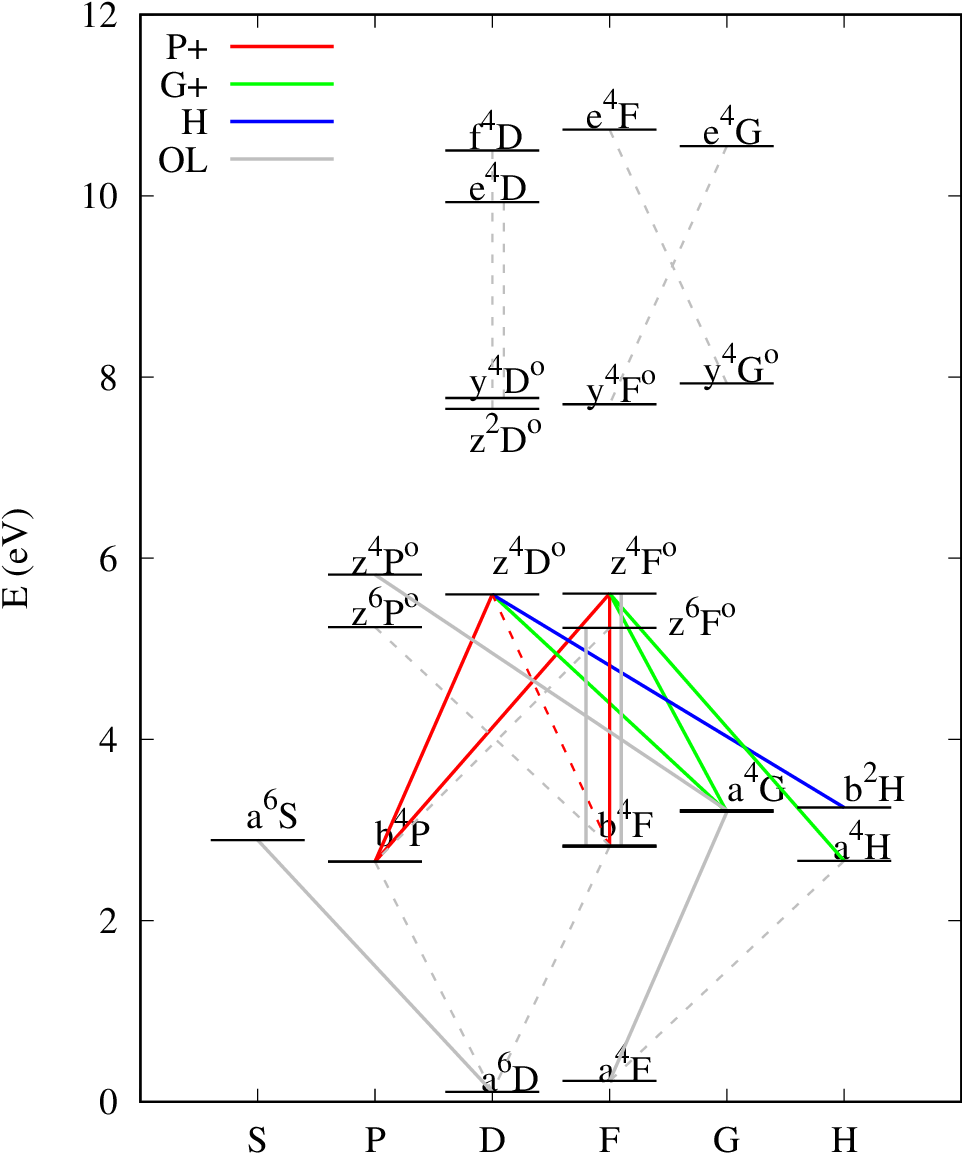}
    \caption{Simplified Grotrian diagram of the inconsistent Fe II lines in the 4000-5600 \AA \ range divided into line groups (P+, G+, H, and OL). The multiplet transitions are denoted with solid lines, while single transitions are denoted with dashed lines.}
    \label{figA1}
    \end{figure}

To empirically measure the relative intensities between inconsistent lines, we chose the spectra of the ten NLSy1 galaxies from our sample with the highest S/N (their S/N in g band is in the range of 35 - 74).\footnote{The NLSy1s which spectra were used are {\tiny SDSS J010226.31-003904.5, SDSS J154732.17+102451.2, SDSS J155909.62+350147.5, SDSS J120719.81+241155.8, SDSS J140621.89+222346.5, SDSS J171304.46+352333.5, SDSS J092247.03+512038.0, SDSS J010535.66-040158.4, SDSS J121549.43+544223.9} and {\tiny SDSS J080131.58+354436.4}.} The average values of measured relative intensities among Fe II lines within each inconsistent line group are adopted, and these values are given in Table \ref{T1}.
The largest lack of Fe II flux in models \cite{Kovacevic2010} and \cite{Shapovalova2012} was in the Fe II lines which belong to the P group and overlap with H$\gamma$ (see comparison between Fe II models given in \citep{Park2022}), which is now improved. The simplified Grotrian diagram of transitions of inconsistent Fe II lines is shown in Fig. \ref{figA1}. The properties of the complex Fe II template are summarised in Table \ref{T2}.

    \clearpage
  \onecolumn{
   
\begin{longtable}{|c | c | c | c| c |c | }
\caption{Consistent and inconsistent Fe II lines in the 4000-5600 \AA \ range \\included in the Fe II template.
\label{T1}}\\
\hline 
Wavelength (\AA) &  Fitting group  & Transition  &  Multiplet & gA ($\cdot 10^3$) &  Relative intensity \\
\hline \hline
\endfirsthead

\caption{continued.}\\
\hline
Wavelength (\AA) &  Fitting group  & Transition  &  Multiplet & gA ($\cdot 10^3$) &  Relative intensity \\
\hline \hline
\endhead
\hline

4489.2& F & z${\ }^4F^o_{5/2}$--b${\ }^4F_{7/2}$  & 37   &360  & 0.110   \\
4508.3& F & z${\ }^4D^o_{1/2}$--b${\ }^4F_{3/2}$  & 38   &1400  & 0.367  \\
4515.4& F & z${\ }^4F^o_{5/2}$--b${\ }^4F_{5/2}$  & 37   &1200  & 0.348  \\
4520.2& F & z${\ }^4F^o_{7/2}$--b${\ }^4F_{9/2}$  & 37   &800  & 0.233   \\
4522.6& F & z${\ }^4D^o_{3/2}$--b${\ }^4F_{5/2}$  & 38   &1600  & 0.859  \\
4534.2& F & z${\ }^4F^o_{5/2}$--b${\ }^4F_{3/2}$  & 37   &120  & 0.029   \\
4541.5& F & z${\ }^4D^o_{3/2}$--b${\ }^4F_{3/2}$  & 38   &360  & 0.078   \\
4549.5& F & z${\ }^4D^o_{5/2}$--b${\ }^4F_{7/2}$  & 38   &6000  & 1.000  \\
4555.9& F & z${\ }^4F^o_{7/2}$--b${\ }^4F_{7/2}$  & 37   &1400  & 0.474  \\
4576.3& F & z${\ }^4D^o_{5/2}$--b${\ }^4F_{5/2}$  & 38   &360  & 0.136   \\
4582.8& F & z${\ }^4F^o_{7/2}$--b${\ }^4F_{5/2}$  & 37   &240  &  0.070  \\
4583.8& F & z${\ }^4D^o_{7/2}$--b${\ }^4F_{9/2}$  & 38   &5600  &  1.353 \\

\hline
   4923.9& S & z${\ }^6P^o_{3/2}$--a${\ }^6S_{5/2}$  & 42   &16000  & 0.693  \\
   5018.4& S & z${\ }^6P^o_{5/2}$--a${\ }^6S_{5/2}$  & 42   &12000  & 1.000  \\
   5169.0& S & z${\ }^6P^o_{7/2}$--a${\ }^6S_{5/2}$  & 42   &32000  & 0.854  \\
   5284.1& S & z${\ }^6F^o_{7/2}$--a${\ }^6S_{5/2}$  &  41* & 160  & 0.019   \\
\hline
 
5197.6 &  G & z${\ }^4F^o_{3/2}$--a${\ }^4G_{5/2}$    &   49     &  2000      & 0.620 \\
5234.6 &  G & z${\ }^4F^o_{5/2}$--a${\ }^4G_{7/2}$    &   49     &  1200      &  0.695\\
5276.0 &  G & z${\ }^4F^o_{7/2}$--a${\ }^4G_{9/2}$    &   49     &  3200      & 0.928 \\
5316.6 &  G & z${\ }^4F^o_{9/2}$--a${\ }^4G_{11/2}$   &   49     &  4000      & 1.000  \\
5316.7 &  G & z${\ }^4D^o_{5/2}$--a${\ }^4G_{7/2}$    &   48*    &  360       & 0.097 \\
5325.5 &  G & z${\ }^4F^o_{7/2}$--a${\ }^4G_{7/2}$    &   49     &  160       & 0.047 \\
                                                           
\hline                                                     

4087.3  &  P+   &  z${\ }^4F^o_{3/2}$--b${\ }^4P_{5/2}$       & 28*     &   12     & 0.118   \\
4122.7  &  P+   &  z${\ }^4F^o_{5/2}$--b${\ }^4P_{5/2}$       & 28*     &   180    & 0.235   \\
4128.7  &  P+   &  z${\ }^4D^o_{3/2}$--b${\ }^4P_{5/2}$       & 27      &   120    & 0.470   \\
4173.5  &  P+   &  z${\ }^4D^o_{5/2}$--b${\ }^4P_{5/2}$       &  27     &   2400   & 1.000   \\                              
4178.8  &  P+   &  z${\ }^4F^o_{7/2}$--b${\ }^4P_{5/2}$       &  28*    &   1600   & 0.245   \\
4233.2  &  P+   &  z${\ }^4D^o_{7/2}$--b${\ }^4P_{5/2}$       &  27     &   5600   & 1.059   \\
4258.1  &  P+   &  z${\ }^4F^o_{3/2}$--b${\ }^4P_{3/2}$       &  28*    &   120   &  0.412    \\
4273.3  &  P+   &  z${\ }^4D^o_{1/2}$--b${\ }^4P_{3/2}$       &  27     &   180   &  0.033    \\
4296.6  &  P+   &  z${\ }^4F^o_{5/2}$--b${\ }^4P_{3/2}$       &  28*   &    420      & 0.765  \\
4303.2  &  P+   &  z${\ }^4D^o_{3/2}$--b${\ }^4P_{3/2}$       &  27    &    800      & 0.470  \\
4351.8  &  P+   &  z${\ }^4D^o_{5/2}$--b${\ }^4P_{3/2}$      &  27    &    3000     &  0.495  \\
4369.4  &  P+   &  z${\ }^4F^o_{3/2}$--b${\ }^4P_{1/2}$      &  28*   &   80       &  0.082   \\
4385.4  &  P+   &  z${\ }^4D^o_{1/2}$--b${\ }^4P_{1/2}$      &  27    &  800      &  0.145     \\
4416.8  &  P+   &  z${\ }^4D^o_{3/2}$--b${\ }^4P_{1/2}$     &   27    &  800      &   0.588  \\
4620.5  &  P+   &  z${\ }^4D^o_{7/2}$--b${\ }^4F_{7/2}$     &   38  &    160         & 0.470     \\
4629.3  &  P+   &  z${\ }^4F^o_{9/2}$--b${\ }^4F_{9/2}$     &   37  &   2000          &  1.059    \\
4666.7  &  P+   &  z${\ }^4F^o_{9/2}$--b${\ }^4F_{7/2}$      &   37  &   100        &   0.706    \\

   \hline  

4278.1  &  G+    &  z${\ }^4F^o_{5/2}$--a${\ }^4H_{7/2}$     &  32*    &   60       &  1.000  \\
4314.3  &  G+    & z${\ }^4F^o_{7/2}$--a${\ }^4H_{9/2}$     &  32*    &   80       & 0.500  \\
4384.3  &   G+   & z${\ }^4F^o_{9/2}$--a${\ }^4H_{11/2}$     &  32*    &   70       &  1.375 \\  
4413.6  &   G+   & z${\ }^4F^o_{9/2}$--a${\ }^4H_{9/2}$     &  32*    &   20       & 0.925  \\  
5264.8  &  G+   &  z${\ }^4D^o_{3/2}$--a${\ }^4G_{3/2}$     &  48*    &  160     &   1.000   \\
5337.7  &  G+  &   z${\ }^4D^o_{5/2}$--a${\ }^4G_{5/2}$    &  48*    &   60      &  0.250    \\
5362.8  &  G+  &   z${\ }^4D^o_{7/2}$--a${\ }^4G_{9/2}$    &  48*    &   600     &   1.375   \\
5414.1  &  G+  &   z${\ }^4D^o_{7/2}$--a${\ }^4G_{7/2}$    &  48*    &   80      &  0.500    \\
5425.2  &  G+  &   z${\ }^4F^o_{9/2}$--a${\ }^4G_{9/2}$    &  49     &   90      &  1.000   \\

  \hline 
 5525.1  &  H &   z${\ }^4D^o_{7/2}$--b${\ }^2H_{9/2}$    &  56]   &    24      &  0.082    \\
 5534.8  & H  &   z${\ }^4F^o_{9/2}$--b${\ }^2H_{11/2}$   &  55]   &    300     &  1.000   \\
   \hline 
 4146.6   & OL  &  a${\ }^4G$--a${\ }^4F $       &   [21F]   & --   &  0.300   \\
 4204.5   &  OL  &  f${\ }^4D$--z${\ }^2D^o  $    &  --  &  --   &  0.200     \\
 4244.0  &  OL   &  a${\ }^4G$--a${\ }^4F $     &   [21F] &  --  & 0.600     \\
 4276.8  &  OL    &   a${\ }^4G$--a${\ }^4F $   &  [21F]  &  --  &  0.320    \\
 4319.6  &  OL    &   a${\ }^4G$--a${\ }^4F $   &   [21F] &   --  &  0.400    \\
 4358.4  &   OL   &  a${\ }^4G$--a${\ }^4F $   &    [21F] &  --   &  0.400    \\
  4403.0  &  OL   &  e${\ }^4G$--y${\ }^4F^o $  &   --     &   --   &  0.300    \\
  4440.8  &   OL   &  e${\ }^4F$--y${\ }^4G^o$  &   --     &  --   &   0.500    \\
  4452.1 &   OL  &   a${\ }^6S$--a${\ }^6D  $      &   [7F]  & --  &  0.800    \\
  4457.9 &   OL  &   b${\ }^4F$--a${\ }^6D  $      &   [6F]  & --  &  0.400    \\
  4472.9 &   OL  &   z${\ }^4F^o_{3/2}$--b${\ }^4F_{5/2}$ &   37    &  80 &  1.400    \\
  4491.4 &   OL  &   z${\ }^4F^o_{3/2}$--b${\ }^4F_{3/2}$ &   37   &  800 & 1.600      \\
  4657.0 &   OL  &   z${\ }^4D^o_{5/2}$--a${\ }^6S_{5/2}$ &   43]   &  60 &  0.600     \\
  4664.4 &   OL  &   b${\ }^4P$--a${\ }^6D$        &   [4F]  &  -- &   0.200    \\
  4670.2 &   OL  &  z${\ }^6F^o_{7/2}$--b${\ }^4P_{5/2}$ &  25]        &  24   &  0.280   \\
  4731.4 &   OL  &  z${\ }^4D^o_{7/2}$--a${\ }^6S_{5/2}$ &  43]        &  240   & 1.000   \\
  4763.8 &   OL  &  z${\ }^4P^o_{5/2}$--a${\ }^4G_{7/2}$ &  50]        &  80   &  0.300   \\
  4780.6 &   OL  &  z${\ }^4P^o_{5/2}$--a${\ }^4G_{5/2}$ &  50]        &  80   &   0.400  \\
  4993.3 &   OL  &  z${\ }^6P^o_{7/2}$--b${\ }^4F_{9/2}$ &  36]          &  80      &  0.200    \\
  5072.4 &   OL  &  a${\ }^4H$--a${\ }^4F$        &  [19F]        &  --      &   0.200    \\
  5100.6 &   OL  &  z${\ }^6F^o_{7/2}$--b${\ }^4F_{9/2}$ &  35]          &   --     &  0.300     \\
  5120.3 &   OL  &  z${\ }^6F^o_{5/2}$--b${\ }^4F_{7/2}$ &  35]          &  10      &  0.200      \\
  5132.7 &   OL  &  z${\ }^6F^o_{9/2}$--b${\ }^4F_{9/2}$ &  35]          &  20      &    0.400    \\
  5136.8 &   OL  &  z${\ }^6F^o_{3/2}$--b${\ }^4F_{5/2}$ &  35]          &  12      &   0.300      \\
  5146.1 &   OL  &  z${\ }^6F^o_{7/2}$--b${\ }^4F_{7/2}$ &  35]          &  20      &   0.200     \\
  5171.6 &   OL  &  z${\ }^6F^o1_{1/2}$--b${\ }^4F_{9/2}$ &  35]          &  10      &   0.200      \\
  5505.2  &   OL  &  e${\ }^4D$--y${\ }^4D^o$      &              &           &    0.400            \\   
        \hline 
        \hline 
        \end{longtable}
 \tablefoot{Multiplet designation: * - intersystem multiplets, ] - intercombination (semi-forbidden) multiplets, [] - forbidden multiplets.}

\begin{table*}[b]
\begin{center}
\caption{Summarised characteristics of the complex Fe II model in 4000-5600 \AA \ range. \label{T2}}
\small
\begin{tabular}{|c | c | c | c| c |c | }
\hline

group name  &  consistent/inconsistent  & transitions   &  double/single Gaussian & rel. intensities &   width \\
 (1) & (2) & (3) & (4) & (5) & (6) \\
\hline 
\hline

\multirow{2}{*}{F} & \multirow{2}{*}{consistent} &  \multirow{2}{*}{allowed} &  \multirow{2}{*}{double} & \multirow{2}{*}{calculated} & \multirow{2}{*}{\scriptsize FWHM Fe II$_{cons}$} \\
 &  &  &  &  &   \\
 \hline
\multirow{2}{*}{S} & \multirow{2}{*}{consistent} &  \multirow{2}{*}{allowed} &  \multirow{2}{*}{double}  & \multirow{2}{*}{calculated} & \multirow{2}{*}{\scriptsize FWHM Fe II$_{cons}$}\\
 &  &  &  &  &   \\
 \hline
\multirow{2}{*}{G} & \multirow{2}{*}{consistent} &  \multirow{2}{*}{allowed} &  \multirow{2}{*}{double}  & \multirow{2}{*}{calculated} & \multirow{2}{*}{\scriptsize FWHM Fe II$_{cons}$}\\
&  &  &  &  &   \\
\hline
\multirow{2}{*}{H} &\multirow{2}{*}{inconsistent} &  \multirow{2}{*}{semi-forbidden} &  \multirow{2}{*}{single}  & \multirow{2}{*}{measured} & \multirow{2}{*}{\scriptsize FWHM Fe II H} \\ 
&  &  &  &  &   \\
\hline

\multirow{2}{*}{P+}&\multirow{2}{*}{inconsistent} &  \multirow{2}{*}{allowed, intersystem} &  \multirow{2}{*}{single}   &\multirow{2}{*}{measured} & \multirow{2}{*}{\scriptsize FWHM Fe II P+ }\\
&  &  &  &  &  \\
\hline
\multirow{2}{*}{G+} &\multirow{2}{*}{inconsistent} &  \multirow{2}{*}{intersystem} &  \multirow{2}{*}{single}   &\multirow{2}{*}{measured} &\multirow{2}{*}{\scriptsize FWHM Fe II G+}\\
&  &  &  &  &   \\
\hline

\multirow{2}{*}{OL} &\multirow{2}{*}{inconsistent} & { \small allowed, semi-forbidden, forbidden, } &  \multirow{2}{*}{single}   &\multirow{2}{*}{measured} &\multirow{2}{*}{\scriptsize FWHM Fe II OL}  \\

&  & { \small lines from high energy levels }&  &  &    \\
\hline

\hline

\end{tabular}
\end{center}
\end{table*}
}

\twocolumn
\section{Comparison of fits with different Fe II templates }\label{C}

\begin{figure}[b] 
\centering
 \includegraphics[width=90mm]{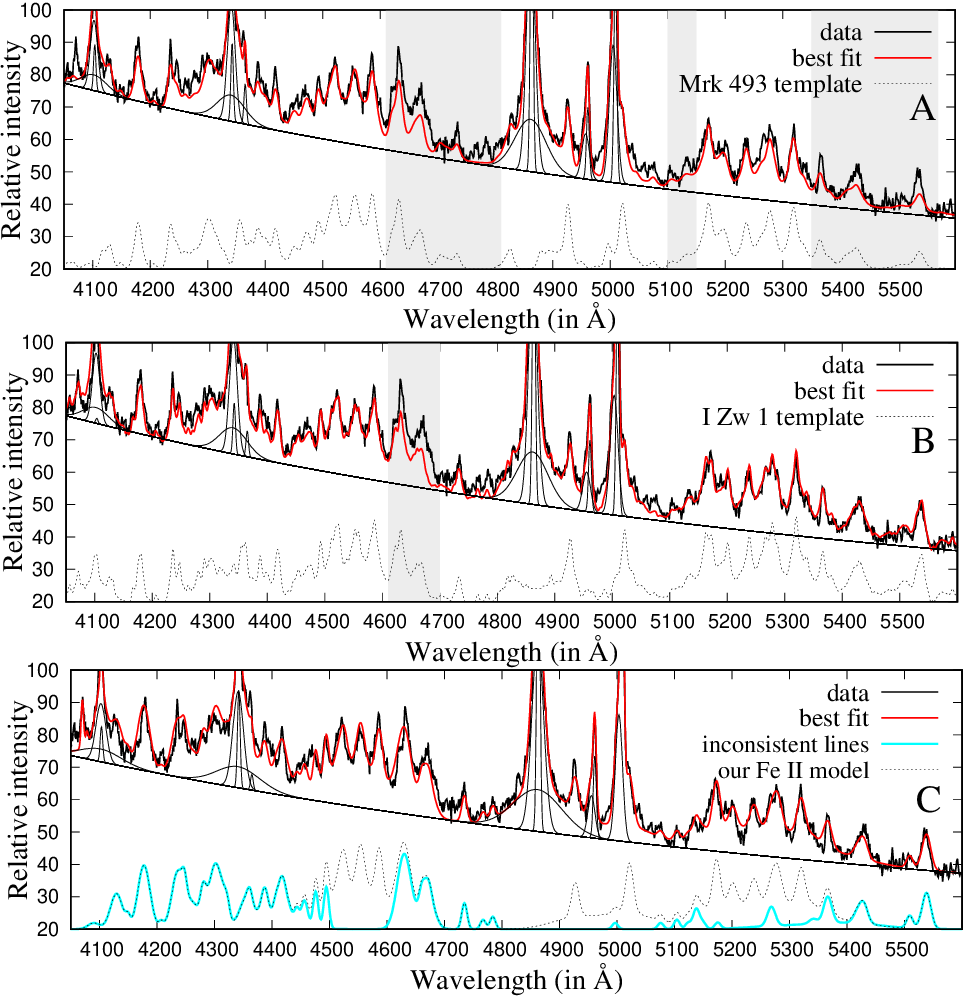}
    \caption{Example of the fit of the spectrum with narrow Fe II lines (SDSS J121549.43+544223.9, plate-mjd-fiber: 8214-57867-0390) with a template made from Mrk 493 given in \cite{Park2022} (see panel A) and with a template made from the I Zw 1 spectrum given in \cite{Boroson1992} (panel B). The shaded regions in panels A and B highlight the mismatch between the fit and data. In panel C we fit the object with our complex Fe II model which assume multiple line groups (F, S and G for consistent and P+, G+, H and OL for inconsistent Fe II lines). Sum of all Fe II lines obtained from best fit with our model is denoted with dashed line, and sum of only inconsistent lines is denoted with cyan.
}
    \label{figCa}
    \end{figure}

\begin{figure}
\centering
 \includegraphics[width=90mm]{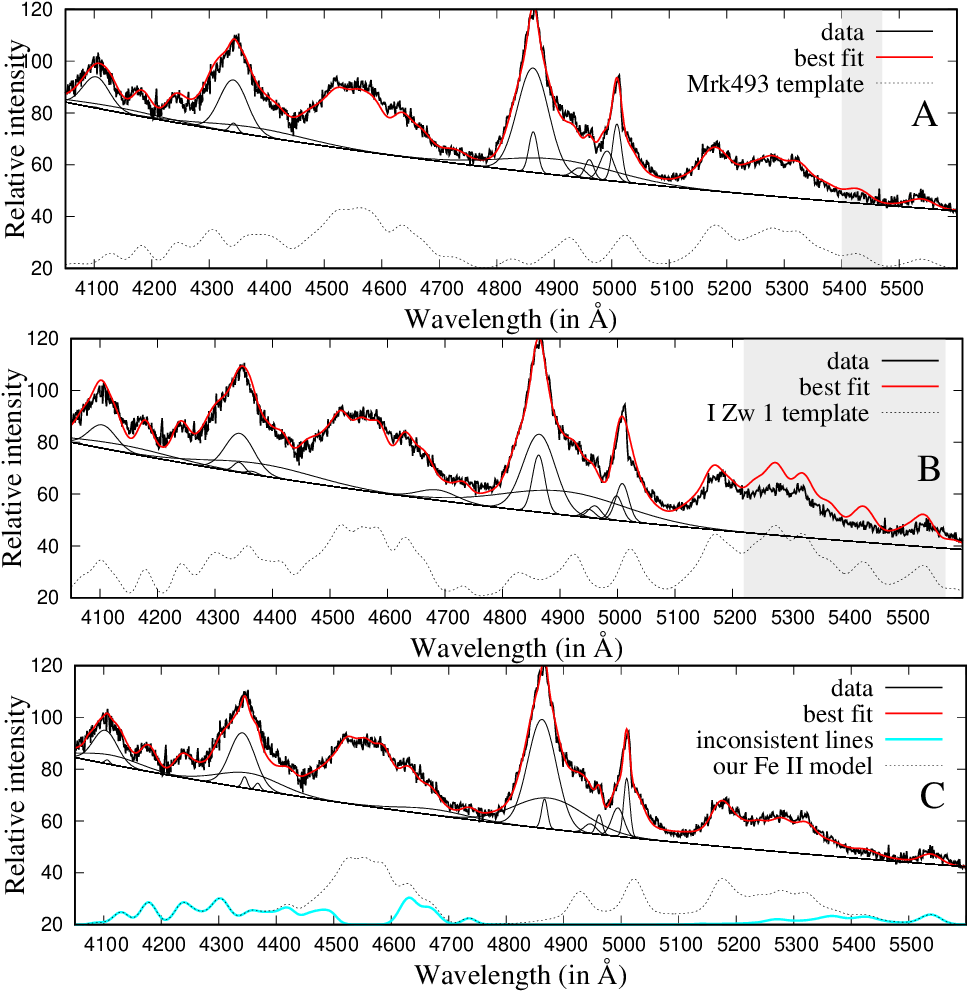}
    \caption{Same as Fig. \ref{figCa} but for the example of a spectrum with broad Fe II lines (SDSS J134251.60-005345.4, plate-mjd-fiber: 0299-51671-0098).}
    \label{figCb}
    \end{figure}

We fitted the spectra with different widths of Fe II lines with two Fe II templates: one made from I Zw 1 spectrum given in \cite{Boroson1992} and the other from Mrk 493 given in \cite{Park2022}. First, we fit one example of spectrum with very narrow  Fe II lines (see Fig. \ref{figCa}), and we found that I Zw 1  template  \citep{Boroson1992} gives slightly better fit than Mrk 493 template  \citep{Park2022}. Both templates give underestimation of the inconsistent lines in range 4600-4700\AA, while in the case of the Mrk 493 template, there is also underestimation of inconsistent lines in red part of Fe II spectrum (see shaded regions in Fig. \ref{figCa}, panel A). However, in the case of the fit of spectrum with broad Fe II lines, Mrk 493 template gives significantly better fit than I Zw 1 template (see Fig. \ref{figCb}). Namely, I Zw 1 template gives overestimation of the inconsistent lines in red part of spectrum (see shaded regions in Fig. \ref{figCb}, panel B). We tried to check can different set of the continuum level give the better fit of this object with I Zw 1 template. We found that when broadening this template to flatten out the inconsistent lines in range 5400-5600 \AA \ and to smear them in pseudocontinuum, the template becomes too broad to fit the Fe II lines bluewards to H$\beta$, specially in 4150-4250 \AA \ range. For each object we gave also the fit with our complex Fe II model (see Figs. \ref{figCa} and \ref{figCb}, panel C), which uses several Fe II line groups (see Sect. \ref{3.1}). We emphasised the inconsistent lines obtained from the best fit with our model with cyan line.

\end{document}